\documentclass[a4paper,12pt]{article}

\usepackage{amsmath}
\usepackage{amssymb}
\usepackage{amsfonts}
\usepackage{amsthm}
\usepackage{bbm}
\usepackage{comment}
\usepackage{color}
\usepackage{cite}
\usepackage{tikz}

\newcommand{\be}{\begin{equation}}
\newcommand{\ee}{\end{equation}}
\newcommand{\bea}{\begin{eqnarray}}
\newcommand{\eea}{\end{eqnarray}}
\newcommand{\bref}[1]{(\ref{#1})}
\newcommand{\nn}{\nonumber}

\newcommand{\mapright}[1]{\smash{[mathop{\hbox to 1cm{\rightarrowfill}}
\limits^{#1}}}

\newcommand{\del}{\partial}

\newcommand{\im}{{\mathbbm i}}
\allowdisplaybreaks[3]

 \makeatletter
 \@addtoreset{equation}{section}
 \makeatother

\usepackage[normalem]{ulem} 

\renewcommand\sout{\bgroup \color{red} \ULdepth=-.5ex \ULset} 

\date{empty}
\begin{document}
\begin{titlepage}
\null
\begin{flushright}
November, 2020
\end{flushright}
\vskip 2cm
\begin{center}
{\Large \bf 
Five-brane Current Algebras in Type II String Theories
}
\vskip 1.5cm
\normalsize
\renewcommand\thefootnote{\alph{footnote}}

{\large
Machiko Hatsuda${}^{*}$\footnote{mhatsuda(at)juntendo.ac.jp}, 
Shin Sasaki${}^{\dagger}$\footnote{shin-s(at)kitasato-u.ac.jp}
and 
Masaya Yata${}^{\ddagger}$\footnote{m-yata(at)juntendo.ac.jp}
}
\vskip 0.5cm
  {\it
  ${}^{*}$
  Department of Radiological Technology, Faculty of Health Science, Juntendo University
  Hongo, Bunkyo-ku, Tokyo 113-0033, Japan \\
  ${}^{*}$KEK Theory Center, High Energy Accelerator Research Organization
  Tsukuba, Ibaraki 305-0801, Japan \\
  \vspace{0.2cm}
  ${}^{\dagger}$
  Department of Physics,  Kitasato University \\
  Sagamihara 252-0373, Japan
\\
  \vspace{0.2cm}
  ${}^{\ddagger}$
  Physics Division, Faculty of Medicine, Juntendo University, Chiba 270-1695, Japan
  }
\vskip 1.5cm
\begin{abstract}
We study the current algebras of the NS5-branes, the Kaluza-Klein (KK) five-branes and
 the exotic $5^2_2$-branes in type IIA/IIB superstring theories.
Their worldvolume theories are governed by the six-dimensional
 $\mathcal{N}= (2,0)$ tensor and the $\mathcal{N} = (1,1)$ vector
 multiplets.
We show that the current algebras are determined through the S- and T-dualities.
The algebras of the $\mathcal{N} = (2,0)$ theories are
characterized by the Dirac bracket caused by the self-dual gauge field
 in the five-brane worldvolumes, while those of the $\mathcal{N} =
 (1,1)$ theories are given by the Poisson bracket.
By the use of these algebras, we examine 
extended spaces in terms of tensor coordinates which are
the representation of ten-dimensional supersymmetry.
We also examine the transition rules of the currents in the type IIA/IIB
 supersymmetry algebras in ten dimensions.
Based on the algebras, we write down the section conditions in the
 extended spaces and gauge transformations of the supergravity fields.
\end{abstract}
\end{center}

\end{titlepage}

\newpage
\setcounter{footnote}{0}
\renewcommand\thefootnote{\arabic{footnote}}
\pagenumbering{arabic}
\tableofcontents

\section{Introduction} \label{sect:introduction}
Dualities are key ingredients to understand string theory and M-theory.
In particular, T-duality is the most fundamental duality in string
theory. This is a distinct feature of the string as an extended object
and it never shows up from a point particle picture.
Indeed, T-duality is interpreted as the canonical transformation in the
worldsheet theory of the fundamental string \cite{Giveon:1988tt}.
There are several formulations of string theory with manifest T-duality.
The doubled space was introduced in \cite{Tseytlin:1990nb}
and double field theory (DFT) \cite{Siegel:1993xq, Siegel:1993th,
Siegel:1993bj, Hull:2009mi}, 
which realizes manifest T-duality with the help of the doubled space, 
has been intensively studied recently.
The doubled space is characterized by the spacetime
(the Kaluza-Klein) coordinate $x^{\mu}$ together with the T-dualized
winding coordinate $\tilde{x}_{\mu}$ \cite{Hohm:2013bwa, Aldazabal:2013sca}.
This is generalized to the U-duality manifest formulation of
supergravity known as exceptional field theory (EFT) based on the exceptional
or extended geometries \cite{Hohm:2013vpa}.
They are extensions of the generalized geometry \cite{Hitchin:2004ut}
that possesses the manifest T-duality.

In addition to the fundamental strings, there are a vast number of extended
objects in string theory.
For example, the D-branes are the most famous extended objects in string theory.
One of the authors studied the current algebras of D$p$-branes in
type II string theories \cite{Hatsuda:2012uk}.
The analysis is generalized to the extended objects in M-theory in
eleven dimensions.
The current algebras of the M2-brane \cite{Hatsuda:2012vm} and the M5-brane 
\cite{Hatsuda:2013dya} are studied in detail.
A remarkable fact about these works is that the U-duality groups such as
$SL(5)$ in seven dimensions and $SO(5,5)$ in six dimensions
are explained by the canonical symmetries in these branes.
Even more, with the Courant brackets on the basis introduced in these algebras, one finds the
structure of the generalized geometries associated with the U(T)-dualities.
This fact indicates that the current algebras of branes dictate the duality
covariant gauge symmetries in the generalized (doubled or exceptional) geometries.

One of the remaining brane family is the NS5-brane and the Kaluza-Klein (KK) 5-brane.
The former is the magnetic dual of the fundamental string while the
latter is obtained by the T-duality transformation of the
NS5-branes along the transverse direction to the brane worldvolume.
They are also indispensable elements in string theory.
Exploiting this procedure further, one finds another kind of branes.
One can continue to perform further T-duality transformations on the
KK5-brane. 
The resulting object is a five-brane of codimension two whose geometry is
characterized by the $O(2,2)$ monodromy \cite{deBoer:2010ud}.
It is known as the exotic $5^2_2$-brane or the Q-brane
\cite{Hassler:2013wsa}. 
Their local geometries are patched together 
not by diffeomorphism or the gauge transformation, but 
the $O(2,2)$ T-duality transformation. 
Therefore the background space of the $5^2_2$-brane loses the conventional meaning of geometries.
This kind of background is known as the T-fold or globally non-geometric space \cite{Hull:2006va}. 
The existence of the exotic branes are necessary in string theory.
Indeed, they appear as the higher dimensional origins of the exotic states in
the U-duality multiplets in lower dimensions \cite{Elitzur:1997zn,
Obers:1998fb}.

The purpose of this paper is to write down the current algebras of the
branes in the remaining cornerstones of string theory.
They include the NS5-branes, the KK5-branes and the exotic $5^2_2$-branes in
type II string theories.
They are related by the several duality transformations and dimensional
reductions of the algebras in the M5-brane and D$p$-branes.
We will explicitly perform these manipulations and complete the
collections of current algebras for brane theories.
It is worthwhile to note that the effective worldvolume theories of
several exotic branes have been constructed in this way
\cite{Chatzistavrakidis:2013jqa, Kimura:2014upa, Blair:2017hhy}.

The organization of this paper is as follows.
In the next section, we introduce the current algebra of the M5-brane in
eleven dimensions. We perform the direct dimensional reduction of the
worldvolume theory of the M5-brane and write down the current algebra of
the type IIA NS5-brane.
We then perform the T-duality transformations of the IIA NS5-brane along the
transverse directions and discuss the algebras of the KK5- and the
exotic $5^2_2$-branes. 
Since the T-duality transformation requires an isometry of geometries,
one of the scalar fields in the brane worldvolume theories
loses its meaning as the geometric fluctuation mode along $x^{\mu}$.
This mode is supplemented, in the T-duality transformation, by another
scalar field associated with the fluctuation along the dual winding
coordinates $\tilde{x}_{\mu}$. We will utilize this fact to write down the algebras of the
KK5- and the exotic $5^2_2$-branes.
These worldvolume theories are governed by the
six-dimensional $\mathcal{N} = (2,0)$ tensor multiplet.
In Section \ref{sect:vector}, we study the current algebras of
five-branes whose effective theories are given by the $\mathcal{N} =
(1,1)$ vector multiplet. 
They stem from that of the D5-brane in type IIB theory.
In Section \ref{sect:SUSY_algebras}, by using the current algebras, 
we study the supersymmetry algebras in ten dimensions. We determine the
central charges in type IIA/IIB supergravities.
Based on the spatial diffeomorphism constraints, we also write down the
section conditions in the extended spaces.
Section \ref{sect:conclusion} is devoted to the conclusion and discussions.

\section{Current algebras in $\mathcal{N} = (2,0)$ theories}
\label{sect:tensor}

In this section, we derive the current algebras of the five-branes whose
worldvolume effective theories are governed by the six-dimensional
$\mathcal{N} = (2,0)$ tensor multiplet
whose bosonic components consist of the five scalar fields and the
self-dual 2-form gauge field.
These include the type IIA NS5-brane, the IIB KK5-brane and the IIA
$5^2_2$-brane in ten dimensions.

We first introduce the current algebra of the M5-brane and then perform the direct dimensional reduction
to obtain the algebra of the type IIA NS5-brane.
The worldvolume effective theory of the M5-brane is governed by the
six-dimensional $\mathcal{N} = (2,0)$ tensor multiplet.
The bosonic sector of the M5-brane is given by the Pasti-Sorokin-Tonin
(PST) action \cite{Pasti:1997gx, Pasti:1996vs}.
The action consists of the Dirac-Born-Infeld (DBI), the self-dual field and
the Wess-Zumino (WZ) parts:
\begin{align}
S = T \int d^6 \sigma
\,
\left(
{\cal L}_{\mbox{ \tiny DBI }} + {\cal L}_{\mbox{ \tiny SD }} + {\cal
 L}_{\mbox{ \tiny WZ }}
\right),
\end{align}
where $T$ is the tension of the M5-brane and each part of the
Lagrangian is defined by 
\begin{align}
{\cal L}_{\mbox{ \tiny DBI }} &= - \sqrt{ -h_{ \tilde{\cal F} } },~~
h_{ \tilde{\cal F} }=\det(h_{ \hat{i} \hat{j} } +  \tilde{F}_{ \hat{i} \hat{j} } ),~~
h_{\hat{i} \hat{j}} = \partial_{ \hat{i} } x^m \partial_{ \hat{j} } x^n \hat{G}_{mn}, ~~
\tilde{\cal F}_{\hat{i} \hat{j}} = h_{\hat{i} \hat{k}} h_{\hat{i} \hat{k}'} \tilde{\cal F}^{ \hat{k} \hat{k}' }, \nonumber\\
{\cal L}_{\mbox{ \tiny SD }} &= {  \sqrt{-h} \over 4 } \tilde{ \cal F}^{ \hat{i} \hat{j} } {\cal F}_{ \hat{i} \hat{j} \hat{k} } n^{\hat{k}},~~
h= \det h_{\hat{i} \hat{j}},~~\hat{i}=0,1,2,{\cdots}5, \nonumber\\
{\cal L}_{\mbox{ \tiny WZ }} &=  \epsilon^{ \hat{i}_1 \cdots \hat{i}_6 } \Bigl( { 1 \over 6! } \hat{C}^{[6]}_{ \hat{i}_1 \cdots \hat{i}_6 } + { 1  \over 2\cdot 3!^2 } { \cal F}_{ \hat{i}_1 \hat{i}_2 \hat{i}_3} \hat{C}^{[3]}_{\hat{i}_4 \hat{i}_5 \hat{j}_6} \Bigr), \nonumber\\
F_{ \hat{i} \hat{j} \hat{k} } &= \partial_{ \hat{i} } A_{ \hat{j} \hat{k} } + \partial_{ \hat{j} } A_{ \hat{k} \hat{i} } + \partial_{ \hat{k} } A_{ \hat{i} \hat{j} },~~~
{\cal F}_{ \hat{i} \hat{j} \hat{k} } = F_{ \hat{i} \hat{j} \hat{k} } -
 \partial_{ \hat{i} }x^m \partial_{ \hat{j} }x^n \partial_{ \hat{k} }x^p
 \hat{C}^{[3]}_{mnp}. 
\label{PST}
\end{align} 
Here $\hat{G}_{mn}$, $\hat{C}^{[3]}$, $\hat{C}^{[6]}$ are the spacetime
metric, the 3-form and its magnetic dual in eleven-dimensional
supergravity. 
$x^m \, (m = 0,1, \ldots, 10)$ are the 
spacetime coordinates represented by 
scalar fields in the six-dimensional worldvolume
while $\sigma^{\hat{i}} \, (\hat{i}= 0,1, \ldots, 5)$ is the worldvolume coordinate.
$F^{[3]}= dA^{[2]}$ is the field strength of the self-dual 2-form
$A^{[2]}$ in six dimensions.
The spacetime fields on the worldvolume are evaluated as the pullback
like $P[\hat{C}^{[3]}]_{\hat{i} \hat{j} \hat{k}} = \hat{C}^{[3]}_{\hat{i} \hat{j} \hat{k}} = \del_{\hat{i}} x^m
\del_{\hat{j}} x^n \del_{\hat{k}} x^p \hat{C}^{[3]}_{mnp}$.
We have also defined the following quantities:
\begin{align}
\tilde{\cal F}^{ \hat{i} \hat{j} } = { 1 \over 3! \sqrt{-h} } \epsilon^{ \hat{i} \hat{j} \hat{k} \hat{k}_1 \hat{k}_2 \hat{k}_3 } n_{ \hat{k} } {\cal  F}_{ \hat{k}_1 \hat{k}_2 \hat{k}_3 },~~~
n_{ \hat{k} } = {  \partial_{ \hat{k} }  a  \over \sqrt{ -h^{ \hat{i} \hat{j} } \partial_{ \hat{i} } a \partial_{\hat{j}} a }},
\end{align}
Here $a$ is the auxiliary field.
The worldvolume indices $\hat{i}, \hat{j}, \ldots$ are 
contracted by the induced metric $h_{\hat{i} \hat{j}} =
P[\hat{G}]_{\hat{i} \hat{j}} = \del_{\hat{i}}
x^m \del_{\hat{j}} x^n \hat{G}_{mn}$ and its inverse $h^{\hat{i} \hat{j}}$.

Based on the PST action, the current algebra of the M5-brane 
was derived \cite{Hatsuda:2013dya}. 
The result is
\begin{align}
&~~~~~\{  Z_M (\sigma), Z_N (\sigma')\}_D = \im \rho^i_{MN} \partial_i \delta(\sigma-\sigma'), \nonumber\\
\rho^i_{MN} &= \begin{bmatrix}
                    0    &   {2 }E^{ji} \partial_j x^{ [n_1 } \delta_m^{n_2]}   &   r^{ i ~:~ n_1 {\cdots} n_5 }_m  \\
                    {2 }E^{ji} \partial_j x^{ [m_1 } \delta_n^{m_2]}  &  r^{ i ~:~ m_1 m_2 n_1 n_2 \ell }_{\ell} &  0  \\
                    r^{ i ~:~ m_1 {\cdots} m_5 }_n  &  0  &  0
                    \end{bmatrix},  \nonumber\\
Z_M =& \begin{bmatrix}
         Z_m \\
         Z^{ [2] m_1 m_2}  \\
         Z^{ [5] m_1 \cdots m_5}
         \end{bmatrix}
       =\begin{bmatrix}
         p_m \\
         2 E^{ i_1 i_2 } \partial_{ i_1 } x^{ m_1 } \partial_{ i_2 } x^{ m_2 } \\
         T \epsilon^{ i_1 \cdots i_5 } \partial_{ i_1 } x^{ m_1 } \partial_{ i_2 } x^{ m_2 } \partial_{ i_3 } x^{ m_3 } \partial_{ i_4 } x^{ m_4 } \partial_{ i_5 } x^{ m_5 }
          \end{bmatrix}, \nonumber\\
E^{ij} =& { 1 \over 4 } \epsilon^{ ij i_1 i_2 i_3 } \partial_{i_1} A_{ i_2 i_3 },~~~
r^{ i ~:~ n_1 \cdots n_5}_m = { 1 \over 4! } \epsilon^{ i i_1 i_2 i_3 i_4 } \partial_{ i_1 } x^{ [n_1 } { \cdots } \partial_{ i_4 } x^{ n_4 } \delta^{ n_5] }_m,
 \label{type0}
\end{align}
where $M,N, \ldots = (m, m_1 m_2, m_1 \ldots m_5)$ are the indices in
the extended space, $\{\cdot, \cdot\}_D$ is the Dirac bracket, 
$p_m$, $E^{ij} =
\frac{1}{4} \epsilon^{ij i_1 i_2 i_3} \del_{i_1} A_{i_2 i_3}$ are the
canonical momenta conjugate to the worldvolume fields $x^m$ and
$A_{ij}$, respectively.
The worldvolume delta function is abbreviated as  $\partial_i\delta(\sigma-\sigma')=\frac{\partial}{\partial \sigma^i}\delta^{5}(\sigma-\sigma')$.
We have defined the spatial indices $i,j, \ldots = 1, 2,
\ldots, 5$ in the worldvolume and $\epsilon^{0 i_1 i_2 \cdots i_5} =
\epsilon^{i_1 i_2 \cdots i_5}$.
The explicit form of the momentum $p_m$ is given by
\begin{align}
p_m 
      &= \tilde{p}_m -  \hat{G}_{mn} t^n + \hat{C}^{[6]}_m - {
 \epsilon^{i_1 \cdots i_5} \over 4!} (2F_{ i_1i_2i_3
 }-\hat{C}^{[3]}_{i_1i_2i_3}) \partial_{i_4}x^{m_1} \partial_{i_5} x^{m_2}
 \hat{C}^{[3]}_{m_1m_2m},
\label{Cmomenta}
\end{align}
where we have defined 
\begin{align}
\tilde{p}_m &= { \partial {\cal L}_{\mbox{ \tiny DBI }} \over
 \partial(\partial_0x^m) }, 
\notag \\
t^n &= \partial_{i_5} x^n { \epsilon^{ i_1 \cdots i_5 } \over 8}
 \tilde{\cal F}_{ i_1i_2 } \tilde{\cal F}_{ i_3i_4 } = \partial_i x^n {
 1 \over 2 } h^{ij} {\cal F}_{ji_1i_2} \tilde{E}^{i_1i_2}, 
\notag \\
\tilde{E}^{ij} &= { 2 \over 4! } \epsilon^{ iji_1i_2i_3 } {\cal F}_{i_1i_2i_3},~~~~~
\hat{C}^{[3]}_{i_1i_2i_3} = \partial_{i_1}x^{n_1} \partial_{i_2}x^{n_2}
 \partial_{i_3}x^{n_3} \hat{C}^{[3]}_{n_1n_2n_3},
\notag \\
\hat{C}^{[6]}_{n} &= { 1 \over 5! } \epsilon^{i_1 \cdots i_5}
 \partial_{i_1}x^{n_1} \partial_{i_2}x^{n_2} \partial_{i_3}x^{n_3}
 \partial_{i_4}x^{n_4} \partial_{i_5}x^{n_5} \hat{C}^{[6]}_{n n_1 n_2
 n_3 n_4 n_5}.
\end{align}
The momentum from the DBI part is explicitly given by
\begin{align}
\tilde{p}_m =& \ 
-{ 1 \over 2 \sqrt{ -h_{ \tilde{\cal F} } } } 
\Bigg[
h'_m 
- \epsilon^{ i_1 \cdots i_5 } 
\Bigl( { 1 \over 6 } \partial_{i_1} x^{m_1} \partial_{i_2} x^{m_2} \partial_{i_3} x^{m_3} \tilde{\cal F}_{ i_4 i_5 } \hat{G}_{ m_1n_1 } \hat{G}_{ m_2n_2 } \hat{G}_{ m_3n_3 } \hat{G}_{ mn_4 } {\cal T }_1^{n_1n_2n_3n_4} \nonumber\\
   &~~~~~~~~~~~~~~~~~~~~~~~~~~~~~~~~~~~~~~~~~~~~~~~~~~~~~~~~~ +{ 1\over8 } \partial_{i_1} x^{m_1} \tilde{\cal F}_{ i_2 i_3 } \tilde{\cal F}_{ i_4 i_5 } \hat{G}_{m_1n_1} \hat{G}_{mn_2} {\cal T }_2^{n_1n_2} \Bigr) \nonumber\\
   &~~~+\epsilon^{ \hat{i}_1 \cdots \hat{i}_6 } \Bigl( {1\over4!}
 \partial_{\hat{i}_1} x^{m_1} \cdots \partial_{\hat{i}_4} x^{m_4} {
 \tilde{\cal F}'_{ m \hat{i}_5 \hat{i}_6 } } \hat{G}_{m_1n_1} \cdots \hat{G}_{m_4n_4} {\cal T }_1^{n_1n_2n_3n_4} \nonumber\\
   &~~~~~~~~~~~~~~~~~~~ + { 1 \over 8 } \partial_{\hat{i}_1} x^{m_1} \partial_{\hat{i}_2} x^{m_2} { \tilde{\cal F}'_{ m \hat{i}_3 \hat{i}_4 } } \tilde{\cal F}_{ \hat{i}_5 \hat{i}_6 } \hat{G}_{m_1n_1} \hat{G}_{m_2n_2} {\cal T }_2^{n_1n_2} 
                             + { 1 \over 4! } { 1 \over 4^2 }
 { \tilde{\cal F}'_{ m \hat{i}_1 \hat{i}_2 } }\tilde{\cal
 F}_{ \hat{i}_3 \hat{i}_4 } \tilde{\cal F}_{ \hat{i}_5 \hat{i}_6 } {\cal
 T}_3 \Bigr)
\Bigg].
\end{align}
Here each quantity is found to be 
\begin{align}
h'_m &= -{ 2 \over 5! } \epsilon^{ i_1 \cdots i_5 } \partial_{i_1} x^{m_1}
 \cdots \partial_{i_5} x^{m_5} \hat{G}_{m_1n_1} \cdots \hat{G}_{m_5n_5} \hat{G}_{mn_6} (
 \epsilon^{ \hat{j}_1 \cdots \hat{j}_6 } \partial_{\hat{i}_1} x^{n_1}
 \cdots \partial_{\hat{i}_6} x^{n_6} ),
\notag \\
\tilde{\cal F}'_{ m \hat{i} \hat{j} }  &= ( \delta_{ 0 \hat{i} }
 \partial_{ \hat{k} } x^p +\delta_{0\hat{k}} \partial_{ \hat{i} } x^p ) \hat{G}_{mp} h_{ \hat{j} \hat{k}' } \tilde{\cal F}^{ \hat{k} \hat{k}' } 
                                              +h_{ \hat{i}\hat{k} } ( \delta_{0\hat{j}} \partial_{ \hat{k}' } x^p + \delta_{0 \hat{k}'} \partial_{\hat{j}} x^p ) \hat{G}_{mp} \tilde{\cal F}^{ \hat{k} \hat{k}' } \nonumber\\
                                          &~~~~~~~+h_{ \hat{i} \hat{k} } h_{ \hat{j} \hat{k}' } \Bigl[ { 1\over3! } {1\over2} (-h)^{-3/2} {h'_{m} } \epsilon^{ \hat{k} \hat{k}' \hat{\ell} \hat{\ell}_1 \hat{\ell}_2 \hat{\ell}_3} n_{\hat{\ell}} {\cal F}_{ \hat{\ell}_1 \hat{\ell}_2 \hat{\ell}_3 } \nonumber\\
                                          &~~~~~~~~~~~~~~~~~~~~~~ - { 1 \over 3! \sqrt{-h} } \epsilon^{\hat{k} \hat{k}' \hat{\ell}  \hat{\ell}_1 \hat{\ell}_2 \hat{\ell}_3 } ( -h^{ \hat{i} \hat{j} } \partial_{\hat{i}} a \partial_{ \hat{j} } a )^{ -3/2 } \partial_{ \hat{\ell} } a \partial_{ \hat{k}_1 }a \partial_{\hat{k}_2} a h^{ \hat{k}_1 0} h^{ \hat{k}_2 \hat{k}_3 } \partial_{\hat{k}_3} x^n 
                                                        \hat{G}_{mn} {\cal F}_{\hat{\ell}_1 \hat{\ell}_2 \hat{\ell}_3} \nonumber\\
                                          &~~~~~~~~~~~~~~~~~~~~~~~+{ 1 \over 2 \sqrt{-h} } \epsilon^{ kk'\ell_1 \ell_2 \ell_2} n_{\ell_1} \partial_{\ell_2} x^p \partial_{\ell_3} x^q \hat{C}^{[3]}_{mpq} \Bigr], \nonumber\\
{\cal T}_1^{m_1 \cdots m_4} &= \epsilon^{ \hat{i}_1 \cdots \hat{i}_6 } \partial_{\hat{i}_1} x^{m_1} \partial_{\hat{i}_2} x^{m_2} \partial_{\hat{i}_3} x^{m_3} \partial_{\hat{i}_4} x^{m_4} \tilde{\cal F}_{ \hat{i}_5 \hat{i}_6 }, \nonumber\\
{\cal T}_2^{m_1 m_2} &= \epsilon^{ \hat{i}_1 \cdots \hat{i}_6 } \partial_{\hat{i}_1} x^{m_1} \partial_{\hat{i}_2} x^{m_2} \tilde{\cal F}_{ \hat{i}_3 \hat{i}_4 } \tilde{\cal F}_{ \hat{i}_5 \hat{i}_6 },~~~~~~~~~
{\cal T}_3 = \epsilon^{ \hat{i}_1 \cdots \hat{i}_6 } \tilde{\cal F}_{ \hat{i}_1 \hat{i}_2 } \tilde{\cal F}_{ \hat{i}_3 \hat{i}_4 } \tilde{\cal F}_{ \hat{i}_5 \hat{i}_6 }.
\end{align}
We note that all the background fields $\hat{G}_{mn}, \hat{C}^{[3]},
\hat{C}^{[6]}$ in eleven-dimensional
supergravity are included in $\tilde{p}_m$ in \bref{Cmomenta}.

For later convenience, we decompose the relation (\ref{type0}) as
\begin{align}
\{ Z_m(\sigma), Z_n(\sigma') \}_D &= 0, \notag \\
\{ Z_m(\sigma), Z^{ [2]n_1n_2 } (\sigma') \}_D &= {2} \im E^{ji}
 \partial_{j} x^{[n_1} \delta^{ n_2] }_m \partial_i
 \delta(\sigma-\sigma'), 
\notag \\
\{ Z_m(\sigma), Z^{ [5]n_1\cdots n_5 } (\sigma') \}_D &= \im  r^{ i~:~
 n_1 \cdots n_5 }_m \partial_i \delta (\sigma -\sigma'), 
\notag \\
\{ Z^{ [2]m_1m_2 }(\sigma), Z^{ [2]n_1n_2 } (\sigma') \}_D &=\im  r^{ i
 ~:~ m_1m_2n_1n_2 \ell }_{\ell} \partial_{i} \delta(\sigma -\sigma'), 
\notag \\
\{ Z^{ [2]m_1m_2 }(\sigma), Z^{ [5]n_1\cdots n_5 } (\sigma') \}_D &= 0,
 \notag \\
\{ Z^{ [5]m_1\cdots m_5 } (\sigma), Z^{ [5]n_1\cdots n_5 } (\sigma')
 \}_D &= 0.
\label{eq:M5_algebra}
\end{align}
We stress that the algebra \eqref{eq:M5_algebra} is characterized by the Dirac bracket.
This stems from the fact that the worldvolume theory of the M5-brane 
is governed by the $\mathcal{N} = (2,0)$ tensor multiplet. 
In which, the self-duality of the 2-form gauge field gives the
second class constraint \cite{Hatsuda:2013dya}.

\subsection{IIA NS5-brane}

We now perform the direct dimensional reduction of the PST action to ten
dimensions and obtain the effective action of the type IIA NS5-brane.
We define $x^{10}=Y$ and perform the dimensional reduction in this direction.
The KK ansatz for the eleven-dimensional metric is 
\begin{align}
\hat{G}_{mn} = 
\begin{bmatrix}
e^{ -{2 \over 3} \phi } ( G_{\mu\nu} + e^{2\phi} C^{[1]}_{\mu} C^{[1]}_{\nu} ) & e^{ { 4 \over 3 } \phi } C^{[1]}_{\mu} \\
e^{ { 4 \over 3 } \phi } C^{[1]}_{\nu}  &  e^{ {4\over3} \phi }
\end{bmatrix}
.
\label{eq:metric_11to10}
\end{align}
Here $G_{\mu \nu}$, $\phi$, $C^{[1]}_{\mu}$ are the spacetime metric,
the dilaton and the RR 1-form in ten-dimensional type IIA supergravity.
The spacetime indices $\mu, \nu, \ldots$ run from 0 to 9.
The potentials in eleven-dimensional supergravity is decomposed as 
\begin{align}
\hat{C}^{[3]}&= C^{[3]} - B \wedge {\rm d} Y, \nonumber\\
\hat{C}^{[6]}&= B^{[6]} + C^{[5]} \wedge {\rm d} Y + {1\over2} C^{[5]}
 \wedge C^{[1]} + {1\over2} C^{[3]} \wedge B \wedge {\rm d} Y, 
\label{eq:forms_11to10}
\end{align}
where $B$ and $B^{[6]}$ are the NSNS $B$-field and its magnetic dual
while $C^{[5]}, C^{[3]}$ are the RR 5- and 3-form potentials in ten dimensions.
Then the PST action of the M5-brane reduces to that of the NS5-brane in
type IIA string theory \cite{Bandos:2000az}:
\begin{align}
 S &= - T_{\text{NS5}} \int d^6 \sigma \, e^{ -2 \phi } 
\sqrt{ - \det ( P[G]_{\hat{i} \hat{j}} + \lambda^2 e^{ 2 \phi }
 F_{\hat{i}} F_{\hat{j}} ) }
\sqrt{  \det  \Bigl( \delta^{\hat{j}}_{\hat{i}} 
+ \im {  \lambda e^{\phi}  (  P[G]_{\hat{i} \hat{k}} 
+ \lambda^2 e^{ 2 \phi } F_{\hat{i}} F_{\hat{k}}  ) \over {\cal N} 
\sqrt{ \det ( \delta_{\hat{l}}^{\hat{m}} + \lambda^2 e^{2\phi}
 F_{\hat{l}} F^{\hat{m}} ) }  
} H^{\ast \hat{j} \hat{k}} \Bigr) } 
\notag \\
&~~~~~ 
- { \lambda^2 \over 4 } T_{\text{NS5}} \int d^6 \sigma \, 
\sqrt{ -G } { 1 \over {\cal N}^2 } 
H^{ \ast \hat{i} \hat{j} } H_{\hat{i} \hat{j} \hat{k}} 
\Bigl( P[G]^{\hat{k} \hat{l}} 
- {  e^{2\phi} \lambda^2 F^{\hat{k}} F^{\hat{l}} \over 1 
+ \lambda^2 e^{2\phi} F^2 } \Bigr) 
{ \partial_{\hat{l}} a \over \sqrt{ (\partial a)^2 } }
\notag \\
&~~~~~~~~~~~~~~~ 
+ \mu_5 \int_{ M_6 } 
\Bigl( P[B^{[6]}]  
+ \lambda P[ C^{[5]} \wedge {\rm d} Y ] 
+ {1\over2} P[ C^{[5]} \wedge C^{[1]} ] 
+ { \lambda \over 2 } P[ C^{[3]} \wedge B \wedge {\rm d} Y  ] 
\notag \\
&~~~~~~~~~~~~~~~~~~~~~~~~~~~~~~~~~~~~~~~~~~~~~~~~~~~~~~~~~~~~~~~~~~~~~~
- { \lambda \over 2 } F^{[3]} \wedge P[C^{[3]}] + {\lambda \over 2}
 F^{[3]} \wedge P[ B \wedge {\rm d} Y ] \Bigr). 
\label{eq:IIANS5_action}
\end{align} 
where $T_{\text{NS5}}$ and $\mu_5$ is the tension and the NSNS charges
of the NS5-brane, $P$ denotes the pullback from the ten-dimensional spacetime to the
six-dimensional worldvolume and $\lambda = 2 \pi \alpha'$ is the string
slope parameter.
We have also defined the following quantities.
\begin{align}
F_{\hat{i}} &= 
\partial_{\hat{i}} Y + \lambda^{-1} P[ C^{[1]}]_{\hat{i}},
~~~
{\cal N} = \Bigl[ 1 - { \lambda^2 e^{2\phi} ( F \partial a )^2 \over  (
 \partial a )^2 ( 1 + \lambda^2 e^{2\phi} F^2 )  } \Bigr]^{1/2},
~~~
G = \det P[G], 
\notag \\
(\partial a)^2 & = P[G]^{\hat{i} \hat{j}} \partial_{\hat{i}} a
 \partial_{\hat{j}} a,
~~~
H_{\hat{i} \hat{j} \hat{k}} = F^{[3]}_{\hat{i} \hat{j} \hat{k}} -
 \lambda^{-1} P[C^{[3]}]_{\hat{i} \hat{j} \hat{k}} - ( P[B] \wedge {\rm
 d} Y )_{\hat{i} \hat{j} \hat{k}}, 
\notag \\
H^{\ast \hat{i} \hat{j} \hat{k}} &= { 1 \over 3! \sqrt{-G} }
 \epsilon^{\hat{i} \hat{j} \hat{k} \hat{m} \hat{n} \hat{p}} H_{\hat{m}
 \hat{n} \hat{p}},
~~~
H^{\ast \hat{i} \hat{j}} = { 1 \over \sqrt{ ( \partial a )^2 } } H^{\ast
 \hat{i} \hat{j} \hat{k}} \partial_{\hat{k}} a.
\end{align}
The action \eqref{eq:IIANS5_action} enables one to write down the
current algebra. This is indeed obtained by the direct dimensional reduction of
 \eqref{eq:M5_algebra}. 
The $Z_M$ in the M5-brane algebra is decomposed as 
\begin{displaymath}
Z_m
\left\{
\begin{array}{l}
Z_{\mu} \\
Z_Y = Z^{[0]}_{\mbox {\tiny RR}}
\end{array}
\right.,~~~
Z^{ [2] m_1 m_2 }
\left\{
\begin{array}{l}
Z^{ [2] \mu \nu } = Z^{[2]\mu\nu}_{\mbox{ \tiny RR }}\\
Z^{ [2] \mu Y } = Z^{ [1]\mu }_{ \mbox{ \tiny NS } }
\end{array}
\right.,~~~
Z^{ [5] m_1 \cdots m_5 }
\left\{
\begin{array}{l}
Z^{ [5] \mu_1 \cdots \mu_5 }\\
Z^{ [5] \mu_1 \cdots \mu_4 Y } = Z^{[4]\mu_1 \cdots \mu_4}_{\mbox{\tiny RR}}
\end{array}
\right.
\end{displaymath}
Here, NS and RR denote 
that $Z$ belongs to the NS- and the RR-sector with each other. 
The rank five anti-symmetric tensors are linear combinations of 
	the self-dual  and anti self-dual tensors,
		where the self-dual and anti self-dual  tensors are
	$Z^{[5\pm]\mu_1\cdots \mu_5}=(Z^{[5]\mu_1\cdots \mu_5}\pm
	\frac{1}{5!}\epsilon^{\mu_1\cdots \mu_{10}}Z^{[5]\mu_6\cdots \mu_{10}})/2$.
	The KK5-brane and the NS5-brane currents are 
	$Z^{[5(+)]}_{\mbox{\tiny KK}}=(Z^{[5+]}+ Z^{[5-]})/2$ 
	and $Z^{[5(-)]}_{\mbox{\tiny NS}}=(Z^{[5+]}- Z^{[5-]})/2$.
Then we find that the non-zero components of the type IIA NS5-brane
algebra are 
\begin{align}
\left\{ Z_{ \mu }(\sigma) , Z^{ [2] \nu \sigma }_{\mbox{\tiny
 RR}}(\sigma')  \right\}_D &= 2 \im E^{ ji } \partial_j  x^{ [\nu }
 \delta^{ \sigma] }_{ \mu } \partial_{i} \delta(\sigma-\sigma'), 
\nonumber \\
\left\{ Z_{ \mu }(\sigma) , Z^{ [1] \nu }_{\mbox{\tiny NS}}(\sigma')
 \right\}_D &=  - 2 \im E^{ ji } \partial_j Y
 \delta^{\nu}_{\mu}  \partial_{i} \delta(\sigma-\sigma'), 
\nonumber \\
\left\{ Z^{ [0] }_{\mbox{\tiny RR}}(\sigma) , Z^{ [1] \nu }_{\mbox{\tiny
 NS}}(\sigma')  \right\}_D &=   2 \im E^{ ji } \partial_j  x^{ \nu }
 \partial_{i} \delta(\sigma-\sigma'), 
\nonumber \\
\left\{ Z_{ \mu }(\sigma) , Z^{ [5] \nu_1 \cdots \nu_5  }(\sigma')
 \right\}_D &= \im \Bigl( { 1 \over 4! } \epsilon^{ i i_1 \cdots i_4 }
 \partial_{ i_1 } x^{ [ \nu_1 } \cdots \partial_{i_4} x^{ \nu_4 }
 \delta^{ \nu_5 ] }_{\mu} \Bigr) \partial_i \delta(\sigma-\sigma' ),
 \nonumber \\
\left\{ Z_{ \mu }(\sigma) , Z^{ [4] \nu_1 \cdots \nu_4 }_{\mbox{\tiny
 RR}}(\sigma')  \right\}_D &=  \im \Bigl( { 1 \over 3! } \epsilon^{ i
 i_1 \cdots i_4 } \partial_{i_1} Y \partial_{i_2} x^{
 [\nu_1 }  \partial_{i_3} x^{ \nu_2 } \partial_{i_4} x^{ \nu_3 }
 \delta^{ \nu_4] }_{\mu} \Bigr) \partial_{i} \delta(\sigma-\sigma'), 
\nonumber \\
\left\{ Z^{ [0] }_{\mbox{\tiny RR}}(\sigma) , Z^{ [4] \nu_1 \cdots \nu_4
 }_{\mbox{\tiny RR}}(\sigma')  \right\}_D &=\im \Bigl( { 1 \over 4! }
 \epsilon^{ i i_1 \cdots i_4 } \partial_{i_1} x^{ [\nu_1 } \cdots
 \partial_{i_4} x^{ \nu_4] } \Bigr) \partial_{i} \delta(\sigma-\sigma'), 
\nonumber \\
\left\{ Z^{ [2] \mu_1 \mu_2 }_{\mbox{\tiny RR}}(\sigma) , Z^{ [2] \nu_1
 \nu_2 }_{\mbox{\tiny RR}}(\sigma')  \right\}_D &= \im  { 7 \over 4! }
 \epsilon^{ i i_1 \cdots i_4 } \partial_{i_1} x^{ [\mu_1 }
 \partial_{i_2} x^{ \mu_2 } \partial_{i_3} x^{ \nu_1 } \partial_{i_4}
 x^{ \nu_2] } \partial_{i} \delta(\sigma-\sigma') ,
\nonumber \\
\left\{ Z^{ [1] \mu_1 }_{\mbox{\tiny NS}}(\sigma) , Z^{ [2] \nu_1 \nu_2
 }_{\mbox{\tiny RR}}(\sigma')  \right\}_D &=  - \im  { 7 \over 6 }
 \epsilon^{ i i_1 \cdots i_4 } \partial_{i_1}  Y
 \partial_{i_2} x^{ [ \mu_1 } \partial_{i_3} x^{ \nu_1 } \partial_{i_4}
 x^{ \nu_2] } \partial_i \delta(\sigma-\sigma'). 
\label{IIA NS5}
\end{align}
Here each component is given by 
\begin{align}
\begin{bmatrix}
Z_{\mu} \\
Z^{[0]}_{\mbox{\tiny RR}} \\
Z^{ [2] \mu_1 \mu_2}_{\mbox{\tiny RR}}  \\
Z^{ [1] \mu}_{\mbox{\tiny NS}} \\
Z^{ [5] \mu_1 \cdots \mu_5} \\
Z^{ [4] \mu_1 \cdots \mu_4 }_{\mbox{\tiny RR}}
\end{bmatrix}
=
\begin{bmatrix}
p_{\mu} \\
p_Y \\
2 E^{ i_1 i_2 } \partial_{ i_1 } x^{ \mu_1 } \partial_{ i_2 } x^{ \mu_2} 
\\
2 E^{ i_1 i_2 } \partial_{ i_1 } x^{ \mu_1 } \partial_{ i_2 } Y 
\\
 \epsilon^{ i_1 \cdots i_5 } 
\partial_{ i_1 } x^{ \mu_1 } \partial_{ i_2 } x^{ \mu_2 } 
\partial_{ i_3 } x^{ \mu_3 } \partial_{ i_4 } x^{ \mu_4 }
\partial_{ i_5 } x^{ \mu_5 }
\\
 \epsilon^{ i_1 \cdots i_5 } 
\partial_{ i_1 } x^{ \mu_1 } \partial_{ i_2 } x^{ \mu_2 } 
\partial_{ i_3 } x^{ \mu_3 } \partial_{ i_4 } x^{ \mu_4 }
\partial_{ i_5 } Y
\end{bmatrix}.
\label{eq:IIANS5_current}
\end{align}
The explicit form of the momenta $p_{\mu}$, $p_Y$, $E^{ij}$ are given
by \eqref{type0} but written by the ten-dimensional quantities
\eqref{eq:metric_11to10} and \eqref{eq:forms_11to10}.
We note that the algebra \eqref{eq:IIANS5_current} is again
characterized by the Dirac bracket.
This is obvious since the worldvolume supermultiplet, the $\mathcal{N} = (2,0)$
tensor multiplet, is inherited from the M5-brane.
The first and the second class constraints coming from the self-dual
property of the 2-form $A^{[2]}$ are taken over to the type IIA
NS5-brane.

\subsection{IIB KK5-brane}
We next determine the current algebra of the type IIB Kaluza-Klein (KK) 5-brane.
This is obtained by the T-duality transformation along a transverse
direction to the type IIA NS5-brane.
The T-duality transformation of backgrounds is given by the famous Buscher rule
\cite{Buscher:1987sk}. 
For example, the T-duality transformation of the spacetime metric and the NSNS
$B$-field along the $x^9$ direction is given by 
\begin{align}
&
G'_{\mu \nu} = G_{\mu \nu} - \frac{G_{9 \mu} G_{9 \nu} - B_{9 \mu} B_{9
 \nu}}{g_{99}},
\qquad 
G'_{9 \mu} = \frac{B_{9 \mu}}{G_{99}},
\qquad
G'_{99} = \frac{1}{G_{99}},
\notag \\
&
B'_{\mu \nu} = B_{\mu \nu} - \frac{G_{9, \mu} B_{9 \nu} - g_{9 \nu} B_{9
 \mu}}{g_{99}},
\qquad 
B'_{9 \mu} = \frac{G_{9 \mu}}{g_{99}},
\label{eq:Buscher_rule}
\end{align}
where $\mu,\nu \not= 9$. The analogous transformations in the RR sector
are also available \cite{Meessen:1998qm}.
The geometry in the transverse directions to the KK5-brane is given by
the Taub-NUT space. 
This is obtained by applying the Buscher rule \eqref{eq:Buscher_rule} to
the NS5-brane background.
The worldvolume effective action of the type IIB KK5-brane is obtained
by applying the Buscher rule to the background of the type IIA NS5-brane
effective action and exchanging the isometry fluctuation $x^9$ to the one in the
dual direction $\tilde{x}^{9}$. 
The latter is nothing but the Lagrange multiplier in deriving the rule
\eqref{eq:Buscher_rule} in the worldsheet theory of the fundamental
strings \cite{Buscher:1987sk}.
For example, this is implemented in worldvolume
theories as the transformation of the pullbacks of the backgrounds:
\begin{align}
P[G]_{\hat{i} \hat{j}} = G_{\mu \nu} \del_{\hat{i}} x^{\mu} \del_{\hat{j}} x^{\nu}
\ \longrightarrow \
&
\sum_{\mu, \nu \not= 9}
\Bigg[
G_{\mu \nu} - \frac{G_{9 \mu} G_{9 \nu} - B_{9 \mu} B_{9 \nu}}{G_{99}}
\Bigg]
\del_{\hat{i}} x^{\mu} \del_{\hat{j}} x^{\nu}
\notag \\
& \ 
+
\sum_{\mu \not= 9}
\frac{B_{9 \mu}}{G_{99}}
\left(
\del_{\hat{i}} \tilde{x}^9 \del_{\hat{j}} x^{\mu} 
+
\del_{\hat{i}} x^{\mu} \del_{\hat{j}} \tilde{x}^9 
\right)
+
\frac{1}{G_{99}}
\del_{\hat{i}} \tilde{x}^9 \del_{\hat{j}} \tilde{x}^9,
\end{align}
Indeed, the worldvolume effective actions of the KK5-branes have been
obtained with this replacement \cite{Eyras:1998hn, Kimura:2014upa}.
It is obvious that the worldvolume effective theory of the IIB KK5-brane is again governed by the
$\mathcal{N} = (2,0)$ tensor multiplet.
Correspondingly, the current algebra of the type IIB KK5-brane is 
again given by the Dirac bracket.
This is written down by applying the T-duality transformation to that of
the IIA NS5-brane.
Now it is straightforward to write down the current algebra of the IIA
KK5-brane. 
We assume that the $x^9$ direction is a transverse direction to the
NS5-brane and perform the T-duality transformation along this direction
in the NS5-brane algebra.
Furthermore, the type IIA currents $Z_M$ is decomposed 
and translated 
into the type IIB currents.
This is given as follows.
The left hand sides are the IIA RR currents and the right hand sides are the IIB RR currents:
\begin{displaymath}
	 Z^{[0]}_{\mbox {\tiny RR}}
	=
	 Z^{[1]9}_{\mbox {\tiny RR}} ,~~~
	Z^{[2] 
}_{\mbox{ \tiny RR }}
	\left\{
	\begin{array}{l}
		Z^{[2]\mu9}_{\mbox{ \tiny RR }} = Z^{[1]\mu}_{\mbox{ \tiny RR }}\\
		Z^{[2]\mu\nu}_{\mbox{ \tiny RR }} = Z^{ [3]\mu\nu9}_{ \mbox{ \tiny RR } }
	\end{array}
	\right.,~~~
	Z^{ [4]
}_{ \mbox{ \tiny RR } }
	\left\{
	\begin{array}{l}
	Z^{ [4] \mu_1 \mu_2 \mu_3 9}_{ \mbox{ \tiny RR } } = Z^{[3] \mu_1\mu_2\mu_3}_{\mbox{ \tiny RR }}\\
	Z^{ [4] \mu_1\cdots\mu_4}_{ \mbox{ \tiny RR } } = Z^{ [5+]\mu_1\cdots\mu_4 9}_{ \mbox{ \tiny RR } }
	\end{array}
	\right.
\end{displaymath}
where $\mu, \nu, \ldots \not= 9$.
As a result, the non-zero components of the IIB KK5-brane algebra is 
\begin{align}
	\left\{ Z_{ \mu }(\sigma) , Z^{ [3] \nu \sigma 9 }_{\mbox{\tiny
			RR}}(\sigma')  \right\}_D &= 2 \im E^{ ji } \partial_j  x^{ [\nu }
	\delta^{ \sigma] }_{ \mu } \partial_{i} \delta(\sigma-\sigma'), 
	\nonumber \\
	\left\{ Z_{ \mu }(\sigma) , Z^{[1] \nu}_{\mbox{\tiny
			RR}}(\sigma')  \right\}_D &= - 2 \im E^{ ji } \partial_j  
	\tilde{x}^{9} 
	\delta^{ \nu }_{ \mu } \partial_{i} \delta(\sigma-\sigma'), 
	\nonumber\\
	\left\{ Z_{9} (\sigma) , Z^{ [1] \nu }_{\mbox{\tiny
			RR}}(\sigma')  \right\}_D &=  
	\left\{ Z^{ [1]9 }_{\mbox{\tiny RR}}(\sigma) , Z^{ [1] \nu }_{\mbox{\tiny
			NS}}(\sigma')  \right\}_D   =   2 \im E^{ ji } \partial_j  x^{ \nu }
	\partial_{i} \delta(\sigma-\sigma'), 
	\nonumber \\
	\left\{ Z_{ 9 }(\sigma) , Z^{ [1] 9 }_{\mbox{\tiny
			NS}}(\sigma')  \right\}_D &= - 2 \im E^{ ji } \partial_j  Y
	\partial_{i} \delta(\sigma-\sigma'), 
	\nonumber \\
	\left\{ Z_{ \mu }(\sigma) , Z^{ [1] \nu }_{\mbox{\tiny NS}}(\sigma')
	\right\}_D &=  - 2 \im E^{ ji } \partial_j Y \delta^{\nu}_{\mu}
	\partial_{i} \delta(\sigma-\sigma'), 
	\nonumber \\
	\left\{ Z^{ [1]9 }_{\mbox{\tiny RR}}(\sigma) , Z^{ [1] 9 
	}_{\mbox{\tiny NS}}(\sigma')  \right\}_D &=   2 \im E^{ ji } \partial_j
	\tilde{x}^{9}  \partial_{i} \delta(\sigma-\sigma'), 
	\nonumber \\
	\left\{ Z_{ \mu }(\sigma) , Z^{ [5] \nu_1 \cdots \nu_5  }(\sigma')
	\right\}_D &= \im  { 1 \over 4! } \epsilon^{ i i_1 \cdots i_4 }
	\partial_{ i_1 } x^{ [ \nu_1 } \cdots \partial_{i_4} x^{ \nu_4 }
	\delta^{ \nu_5 ] }_{\mu}  \partial_i \delta(\sigma-\sigma' ),  
	\nonumber \\
	\left\{ Z_{ \mu }(\sigma) , Z^{ [5] \nu_1 \cdots \nu_4 9
	}(\sigma')  \right\}_D &=  \im  { 1 \over 3! } \epsilon^{ i i_1
		\cdots i_4 } \partial_{i_1} \tilde{x}^{9} \partial_{i_2} x^{ [\nu_1 }
	\partial_{i_3} x^{ \nu_2 } \partial_{i_4} x^{ \nu_3 } \delta^{ \nu_4]
	}_{\mu}  \partial_{i} \delta(\sigma-\sigma'), 
	\nonumber \\
	\left\{ Z_{ 9 }(\sigma) , Z^{ [5] \nu_1 \cdots \nu_4 9
	}(\sigma')  \right\}_D &=  
	\left\{ Z^{ [1] 9}_{\mbox{\tiny RR}}(\sigma) , Z^{ [5+] \nu_1 \cdots \nu_4 9
	}_{\mbox{\tiny RR}}(\sigma')  \right\}_D 
	\nonumber\\
	&=
	\im { 1 \over 4! }
	\epsilon^{ i i_1 \cdots i_4 } \partial_{i_1} x^{ [\nu_1 } \cdots
	\partial_{i_4} x^{ \nu_4] }  \partial_{i} \delta(\sigma-\sigma'),
	\nonumber \\ 
	\left\{ Z_{ \mu }(\sigma) , Z^{ [5+] \nu_1 \cdots \nu_4 9 }_{\mbox{\tiny
			RR}}(\sigma')  \right\}_D &=  \im  { 1 \over 3! } \epsilon^{ i i_1
		\cdots i_4 } \partial_{i_1} Y \partial_{i_2} x^{ [\nu_1 }
	\partial_{i_3} x^{ \nu_2 } \partial_{i_4} x^{ \nu_3 } \delta^{ \nu_4]
	}_{\mu}  \partial_{i} \delta(\sigma-\sigma'), 
	\nonumber \\
	\left\{ Z_{ 9 }(\sigma) , Z^{ [3] \nu_1 \nu_2 \nu_3
	}_{\mbox{\tiny RR}}(\sigma')  \right\}_D &=  \im  { 1 \over 3! }
	\epsilon^{ i i_1 \cdots i_4 } \partial_{i_1} Y \partial_{i_2} x^{
		[\nu_1 }  \partial_{i_3} x^{ \nu_2 } \partial_{i_4} x^{ \nu_3] }
	\partial_{i} \delta(\sigma-\sigma'), 
	\nonumber \\
	\left\{ Z_{ \mu }(\sigma) , Z^{ [3] \nu_1 \nu_2 \nu_3
	}_{\mbox{\tiny RR}}(\sigma')  \right\}_D &=  -\im  { 1 \over 2! }
	\epsilon^{ i i_1 \cdots i_4 } \partial_{i_1} Y \partial_{i_2} \tilde{x}^{
		9 }  \partial_{i_3} x^{ [\nu_1 } \partial_{i_4} x^{ \nu_2 }
	\delta^{ \nu_3] }_{\mu}  \partial_{i} \delta(\sigma-\sigma'), 
	\nonumber \\
	\left\{ Z^{ [1]9 }_{\mbox{\tiny RR}}(\sigma) , Z^{
		[ 3] \nu_1 \nu_2 \nu_3  }_{\mbox{\tiny RR}}(\sigma')
	\right\}_D &=\im { 1 \over 3! } \epsilon^{ i i_1 \cdots i_4 }
	\partial_{i_1} x^{ [\nu_1 } \partial_{i_2} x^{ \nu_2} \partial_{i_3}
	x^{ \nu_3] } \partial_{i_4} \tilde{x}^{9} \partial_{i}
	\delta(\sigma-\sigma'), 
	\nonumber \\
	\left\{ Z^{ [3] \mu_1 \mu_2 9 }_{\mbox{\tiny RR}}(\sigma) , Z^{ [ 3] \nu_1
		\nu_2 9 }_{\mbox{\tiny RR}}(\sigma')  \right\}_D &= \im { 7 \over 4! }
	\epsilon^{ i i_1 \cdots i_4 } \partial_{i_1} x^{ [\mu_1 }
	\partial_{i_2} x^{ \mu_2 } \partial_{i_3} x^{ \nu_1 } \partial_{i_4}
	x^{ \nu_2] } \partial_{i} \delta(\sigma-\sigma'), 
	\nonumber \\
	\left\{ Z^{ [1] \mu_1 }_{\mbox{\tiny RR}}(\sigma) , Z^{ [3]
		\nu_1 \nu_2 9}_{\mbox{\tiny RR}}(\sigma')  \right\}_D &=  - \im  { 7
		\over 6 } \epsilon^{ i i_1 \cdots i_4 } \partial_{i_1} \tilde{x}^{9}
	\partial_{i_2} x^{ [ \mu_1 } \partial_{i_3} x^{ \nu_1 } \partial_{i_4}
	x^{ \nu_2] } \partial_i \delta(\sigma-\sigma'), 
	\nonumber \\
	\left\{ Z^{ [1] \mu_1 }_{\mbox{\tiny NS}}(\sigma) , Z^{ [ 3] \nu_1 \nu_2 9
	}_{\mbox{\tiny RR}}(\sigma')  \right\}_D &=  - \im  { 7 \over 6 }
	\epsilon^{ i i_1 \cdots i_4 } \partial_{i_1} Y \partial_{i_2} x^{ [
		\mu_1 } \partial_{i_3} x^{ \nu_1 } \partial_{i_4} x^{ \nu_2] }
	\partial_i \delta(\sigma-\sigma'), 
	\nonumber \\
	\left\{ Z^{ [1] \mu_1 }_{\mbox{\tiny NS}}(\sigma) , Z^{ [1] \nu_1
		 }_{\mbox{\tiny RR}}(\sigma')  \right\}_D &=  - \im  { 7
		\over 2 } \epsilon^{ i i_1 \cdots i_4 } \partial_{i_1} Y \partial_{i_2}
	\tilde{x}^{9 } \partial_{i_3} x^{ [ \mu_1 } \partial_{i_4} x^{ \nu_1]
	} \partial_i \delta(\sigma-\sigma'), 
	\nonumber \\
	\left\{ Z^{ [1] 9 }_{\mbox{\tiny
			NS}}(\sigma) , Z^{ [3] \nu_1 \nu_2 9}_{\mbox{\tiny RR}}(\sigma')
	\right\}_D &=  - \im  { 7 \over 2 } \epsilon^{ i i_1 \cdots i_4 }
	\partial_{i_1}  Y \partial_{i_2} \tilde{x}^{ 9 } \partial_{i_3} x^{
		[\nu_1 } \partial_{i_4} x^{ \nu_2] } \partial_i
	\delta(\sigma-\sigma'). 
	\label{KKMcurrent}
\end{align}
Here $\mu, \nu, \ldots \not= 9$ and 
we have shown only the non-zero contributions.
Each component is given by \eqref{eq:IIANS5_current} but all the
background fields in the momenta $p_{\mu} (\mu \not= 9), p_9, p_Y$ are replaced by
the T-dualized ones. 
We note that the KK5-brane possesses a particular transverse direction
that corresponds to the isometry in the Taub-NUT space.
Reflecting this fact, we have a distinguished scalar field $\tilde{x}^9$
which contrasts the algebra \eqref{KKMcurrent} with \eqref{IIA NS5}.

\subsection{IIA $5^2_2$-brane}
We apply the second T-duality transformation in the type IIB KK5-brane along
another transverse, say, the $x^8$-direction.
The resulting object is known as the $5^2_2$-brane or the exotic Q-brane or
the T-fold.
It is an object of the codimension two.
Although it loses the role of the consistent global solution as a stand alone
object, it is nevertheless allowed as a local solution in supergravity \cite{deBoer:2010ud}.
The background fields of the $5^2_2$-brane have a peculiar structure,
namely, they are not single valued functions of spacetime but are patched
together by the non-trivial $O(2,2)$ monodromy.
In this sense, the background space of the $5^2_2$-brane ceases to be a
conventional geometry and the brane is called the (globally)
non-geometric object.

The worldvolume effective action of the IIB $5^2_2$-brane is obtained by the
same procedure discussed in the previous subsection \cite{Kimura:2014upa}.
The effective theory is again characterized by the six-dimensional
$\mathcal{N} = (2,0)$ tensor multiplet where the scalar field $x^8$ in
the isometry direction is replaced by its dual $\tilde{x}^8$.
We find that the non-zero components of the current algebra in type IIA $5^2_2$-brane
are 
\begin{align}
\left\{ Z_{ \mu }(\sigma) , Z^{ [2] \nu \sigma }_{\mbox{\tiny
 RR}}(\sigma')  \right\}_D &= 2 \im E^{ ji } \partial_j  x^{ [\nu }
 \delta^{ \sigma] }_{ \mu } \partial_{i} \delta(\sigma-\sigma'), 
\nonumber \\
\left\{ Z_{ \mu }(\sigma) , Z^{ [2] \nu 8 }_{\mbox{\tiny
 RR}}(\sigma')  \right\}_D &= - 2 \im E^{ ji } \partial_j  \tilde{x}^{8
 } \delta^{ \nu }_{ \mu } \partial_{i} \delta(\sigma-\sigma'), 
\nonumber \\
\left\{ Z_{ \mu }(\sigma) , Z^{ [2] \nu 9 }_{\mbox{\tiny
 RR}}(\sigma')  \right\}_D &= - 2 \im E^{ ji } \partial_j  \tilde{x}^{9}
 \delta^{ \nu }_{ \mu } \partial_{i} \delta(\sigma-\sigma'), 
\nonumber \\
\left\{ Z_{ 9 }(\sigma) , Z^{ [2] \nu 9 }_{\mbox{\tiny
 RR}}(\sigma')  \right\}_D &= 
\left\{ Z_{8}(\sigma) , Z^{ [2] \nu 8 }_{\mbox{\tiny
 RR}}(\sigma')  \right\}_D = 
\left\{ Z^{ [0] }_{\mbox{\tiny RR}}(\sigma) , Z^{ [1] \nu }_{\mbox{\tiny
 NS}}(\sigma')  \right\}_D \nonumber\\
&=   2 \im E^{ ji } \partial_j  x^{ \nu }
 \partial_{i} \delta(\sigma-\sigma'), 
\nonumber \\
\left\{ Z_{9}(\sigma) , Z^{ [2] 8 9
 }_{\mbox{\tiny RR}}(\sigma')  \right\}_D &= 
\left\{ Z^{ [0] }_{\mbox{\tiny RR}}(\sigma) , Z^{ [1] 8
 }_{\mbox{\tiny NS}}(\sigma')  \right\}_D =   2 \im E^{ ji } \partial_j
 \tilde{x}^{8}  \partial_{i} \delta(\sigma-\sigma'), 
\nonumber \\
\left\{ Z_{8}(\sigma) , Z^{ [2] 98
 }_{\mbox{\tiny RR}}(\sigma')  \right\}_D &= 
\left\{ Z^{ [0] }_{\mbox{\tiny RR}}(\sigma) , Z^{ [1] 9 }_{\mbox{\tiny
 NS}}(\sigma')  \right\}_D =   2 \im E^{ ji } \partial_j
 \tilde{x}^{9}  \partial_{i} \delta(\sigma-\sigma'), 
\nonumber \\
\left\{ Z_{9}(\sigma) , Z^{ [1] 9 }_{\mbox{\tiny
 NS}}(\sigma')  \right\}_D &= 
\left\{ Z_{8}(\sigma) , Z^{ [1] 8}_{\mbox{\tiny
 NS}}(\sigma')  \right\}_D = - 2 \im E^{ ji } \partial_j  Y
 \partial_{i} \delta(\sigma-\sigma'), 
\nonumber \\
\left\{ Z_{ \mu }(\sigma) , Z^{ [1] \nu }_{\mbox{\tiny NS}}(\sigma')
 \right\}_D &=  - 2 \im E^{ ji } \partial_j Y \delta^{\nu}_{\mu}
 \partial_{i} \delta(\sigma-\sigma'), 
\nonumber \\
\left\{ Z_{ \mu }(\sigma) , Z^{ [5] \nu_1 \cdots \nu_5  }(\sigma')
 \right\}_D &= \im  { 1 \over 4! } \epsilon^{ i i_1 \cdots i_4 }
 \partial_{ i_1 } x^{ [ \nu_1 } \cdots \partial_{i_4} x^{ \nu_4 }
 \delta^{ \nu_5 ] }_{\mu}  \partial_i \delta(\sigma-\sigma' ),   
\nonumber \\
\left\{ Z_{ \mu }(\sigma) , Z^{ [5] \nu_1 \cdots \nu_4 8
 }(\sigma')  \right\}_D &=  \im  { 1 \over 3! } \epsilon^{ i i_1
 \cdots i_4 } \partial_{i_1} \tilde{x}^{8} \partial_{i_2} x^{ [\nu_1 }
 \partial_{i_3} x^{ \nu_2 } \partial_{i_4} x^{ \nu_3 } \delta^{ \nu_4]
 }_{\mu}  \partial_{i} \delta(\sigma-\sigma'), 
\nonumber \\
\left\{ Z_{ \mu }(\sigma) , Z^{ [5] \nu_1 \cdots \nu_4 9
 }(\sigma')  \right\}_D &=  \im  { 1 \over 3! } \epsilon^{ i i_1
 \cdots i_4 } \partial_{i_1} \tilde{x}^{9} \partial_{i_2} x^{ [\nu_1 }
 \partial_{i_3} x^{ \nu_2 } \partial_{i_4} x^{ \nu_3 } \delta^{ \nu_4]
 }_{\mu}  \partial_{i} \delta(\sigma-\sigma'), 
\nonumber \\
\left\{ Z_{ \mu }(\sigma) , Z^{ [5] \nu_1 \nu_2 \nu_3 9
 8 }(\sigma')  \right\}_D &=  -\im  { 1 \over 2! } \epsilon^{
 i i_1 \cdots i_4 } \partial_{i_1} \tilde{x}^8 \partial_{i_2} \tilde{x}^{
 9 }  \partial_{i_3} x^{ [\nu_1 } \partial_{i_4} x^{ \nu_2 }
 \delta^{ \nu_3] }_{\mu}  \partial_{i} \delta(\sigma-\sigma'), 
\nonumber \\
\left\{ Z_{ 9 }(\sigma) , Z^{ [5] \nu_1 \cdots \nu_4 9
 }(\sigma')  \right\}_D &= 
\left\{ Z_{ 8 }(\sigma) , Z^{ [5] \nu_1 \cdots \nu_4 8
 }(\sigma')  \right\}_D = 
\left\{ Z^{ [0] }_{\mbox{\tiny RR}}(\sigma) , Z^{ [4] \nu_1 \cdots \nu_4
 }_{\mbox{\tiny RR}}(\sigma')  \right\}_D \nonumber\\
&=\im { 1 \over 4! }
 \epsilon^{ i i_1 \cdots i_4 } \partial_{i_1} x^{ [\nu_1 } \cdots
 \partial_{i_4} x^{ \nu_4] }  \partial_{i} \delta(\sigma-\sigma'), 
\nonumber \\
\left\{ Z_{ 9 }(\sigma) , Z^{ [5] \nu_1 \nu_2 \nu_3 8
 9 }(\sigma')  \right\}_D &=  
\left\{ Z^{ [0] }_{\mbox{\tiny RR}}(\sigma) , Z^{ [4] \nu_1 \nu_2 \nu_3
 8 }_{\mbox{\tiny RR}}(\sigma')  \right\}_D \nonumber\\
&=\im { 1 \over 3!
 } \epsilon^{ i i_1 \cdots i_4 } \partial_{i_1} x^{ [\nu_1 }
 \partial_{i_2} x^{ \nu_2} \partial_{i_3} x^{ \nu_3] } \partial_{i_4}
 \tilde{x}^{ 8} \partial_{i} \delta(\sigma-\sigma'), 
\nonumber \\
\left\{ Z_{ 8 }(\sigma) , Z^{ [5] \nu_1 \nu_2 \nu_3 9
 8 }(\sigma')  \right\}_D &=  
\left\{ Z^{ [0] }_{\mbox{\tiny RR}}(\sigma) , Z^{ [4] \nu_1 \nu_2 \nu_3
 9 }_{\mbox{\tiny RR}}(\sigma')  \right\}_D \nonumber\\
&=\im { 1 \over 3!
 } \epsilon^{ i i_1 \cdots i_4 } \partial_{i_1} x^{ [\nu_1 }
 \partial_{i_2} x^{ \nu_2} \partial_{i_3} x^{ \nu_3] } \partial_{i_4}
 \tilde{x}^{ 9} \partial_{i} \delta(\sigma-\sigma'), 
\nonumber \\
\left\{ Z_{ \mu }(\sigma) , Z^{ [4] \nu_1 \cdots \nu_4 }_{\mbox{\tiny
 RR}}(\sigma')  \right\}_D &=  \im  { 1 \over 3! } \epsilon^{ i i_1
 \cdots i_4 } \partial_{i_1} Y \partial_{i_2} x^{ [\nu_1 }
 \partial_{i_3} x^{ \nu_2 } \partial_{i_4} x^{ \nu_3 } \delta^{ \nu_4]
 }_{\mu}  \partial_{i} \delta(\sigma-\sigma'), 
\nonumber\\
\left\{ Z_{ \mu }(\sigma) , Z^{ [4] \nu_1 \nu_2 \nu_3 8
 }_{\mbox{\tiny RR}}(\sigma')  \right\}_D &=  -\im  { 1 \over 2! }
 \epsilon^{ i i_1 \cdots i_4 } \partial_{i_1} Y \partial_{i_2} \tilde{x}^{
 8 }  \partial_{i_3} x^{ [\nu_1 } \partial_{i_4} x^{ \nu_2 }
 \delta^{ \nu_3] }_{\mu}  \partial_{i} \delta(\sigma-\sigma'), 
\nonumber \\
\left\{ Z_{ \mu }(\sigma) , Z^{ [4] \nu_1 \nu_2 \nu_3 9
 }_{\mbox{\tiny RR}}(\sigma')  \right\}_D &=  -\im  { 1 \over 2! }
 \epsilon^{ i i_1 \cdots i_4 } \partial_{i_1} Y \partial_{i_2} \tilde{x}^{
 9 }  \partial_{i_3} x^{ [\nu_1 } \partial_{i_4} x^{ \nu_2 }
 \delta^{ \nu_3] }_{\mu}  \partial_{i} \delta(\sigma-\sigma'), 
\nonumber \\
\left\{ Z_{ \mu }(\sigma) , Z^{ [4] \nu_1 \nu_2 8 9
 }_{\mbox{\tiny RR}}(\sigma')  \right\}_D &=  -\im  \epsilon^{ i i_1
 \cdots i_4 } \partial_{i_1} Y \partial_{i_2} \tilde{x}^{ 9 }
 \partial_{i_3} \tilde{x}^{ 8 } \partial_{i_4} x^{ [\nu_1 } \delta^{
 \nu_2] }_{\mu}  \partial_{i} \delta(\sigma-\sigma'), 
\nonumber \\
\left\{ Z_{ 8 }(\sigma) , Z^{ [4] \nu_1 \nu_2 \nu_3 8
 }_{\mbox{\tiny RR}}(\sigma')  \right\}_D &=  
\left\{ Z_{ 9 }(\sigma) , Z^{ [4] \nu_1 \nu_2 \nu_3 9
 }_{\mbox{\tiny RR}}(\sigma')  \right\}_D \nonumber\\
&=  \im  { 1 \over 3! }
 \epsilon^{ i i_1 \cdots i_4 } \partial_{i_1} Y \partial_{i_2} x^{
 [\nu_1 }  \partial_{i_3} x^{ \nu_2 } \partial_{i_4} x^{ \nu_3] }
 \partial_{i} \delta(\sigma-\sigma'), 
\nonumber \\
\left\{ Z_{ 8 }(\sigma) , Z^{ [4] \nu_1 \nu_2 9
 8 }_{\mbox{\tiny RR}}(\sigma')  \right\}_D &=  \im  { 1 \over
 2! } \epsilon^{ i i_1 \cdots i_4 } \partial_{i_1} Y \partial_{i_2} \tilde{x}^{
 9 }  \partial_{i_3} x^{ [\nu_1 } \partial_{i_4} x^{ \nu_2] }
 \partial_{i} \delta(\sigma-\sigma'), 
\nonumber \\
\left\{ Z_{ 9 }(\sigma) , Z^{ [4] \nu_1 \nu_2 8
 9 }_{\mbox{\tiny RR}}(\sigma')  \right\}_D &=  \im  { 1 \over
 2! } \epsilon^{ i i_1 \cdots i_4 } \partial_{i_1} Y \partial_{i_2} \tilde{x}^{
 8 }  \partial_{i_3} x^{ [\nu_1 } \partial_{i_4} x^{ \nu_2] }
 \partial_{i} \delta(\sigma-\sigma'), 
\nonumber \\
\left\{ Z^{ [0] }_{\mbox{\tiny RR}}(\sigma) , Z^{ [4] \nu_1 \nu_2
 8 9 }_{\mbox{\tiny RR}}(\sigma')  \right\}_D &=\im {
 1 \over 2! } \epsilon^{ i i_1 \cdots i_4 } \partial_{i_1} x^{ [\nu_1 }
 \partial_{i_2} x^{ \nu_2]} \partial_{i_3} \tilde{x}^{ 8 }
 \partial_{i_4} \tilde{x}^{ 9} \partial_{i} \delta(\sigma-\sigma'), 
\nonumber \\
\left\{ Z^{ [2] \mu_1 \mu_2 }_{\mbox{\tiny RR}}(\sigma) , Z^{ [2] \nu_1
 \nu_2 }_{\mbox{\tiny RR}}(\sigma')  \right\}_D &= \im  { 7 \over 4! }
 \epsilon^{ i i_1 \cdots i_4 } \partial_{i_1} x^{ [\mu_1 }
 \partial_{i_2} x^{ \mu_2 } \partial_{i_3} x^{ \nu_1 } \partial_{i_4}
 x^{ \nu_2] } \partial_{i} \delta(\sigma-\sigma') , 
\nonumber \\
\left\{ Z^{ [2] \mu_1 8 }_{\mbox{\tiny RR}}(\sigma) , Z^{ [2]
 \nu_1 \nu_2 }_{\mbox{\tiny RR}}(\sigma')  \right\}_D &=  - \im  { 7
 \over 6 } \epsilon^{ i i_1 \cdots i_4 } \partial_{i_1} \tilde{x}^8
 \partial_{i_2} x^{ [ \mu_1 } \partial_{i_3} x^{ \nu_1 } \partial_{i_4}
 x^{ \nu_2] } \partial_i \delta(\sigma-\sigma'), 
\nonumber \\
\left\{ Z^{ [2] \mu_1 9 }_{\mbox{\tiny RR}}(\sigma) , Z^{ [2]
 \nu_1 \nu_2 }_{\mbox{\tiny RR}}(\sigma')  \right\}_D &=  - \im  { 7
 \over 6 } \epsilon^{ i i_1 \cdots i_4 } \partial_{i_1} \tilde{x}^9
 \partial_{i_2} x^{ [ \mu_1 } \partial_{i_3} x^{ \nu_1 } \partial_{i_4}
 x^{ \nu_2] } \partial_i \delta(\sigma-\sigma'), 
\nonumber \\
\left\{ Z^{ [2] 8 9 }_{\mbox{\tiny RR}}(\sigma) , Z^{
 [2] \nu_1 \nu_2 }_{\mbox{\tiny RR}}(\sigma')  \right\}_D &=  - \im  {
 7 \over 2 } \epsilon^{ i i_1 \cdots i_4 } \partial_{i_1} \tilde{x}^9
 \partial_{i_2} \tilde{x}^{  8 } \partial_{i_3} x^{ [\nu_1 }
 \partial_{i_4} x^{ \nu_2] } \partial_i \delta(\sigma-\sigma'), 
\nonumber \\
\left\{ Z^{ [2] \mu_1 8 }_{\mbox{\tiny RR}}(\sigma) , Z^{ [2]
 \nu_1 9 }_{\mbox{\tiny RR}}(\sigma')  \right\}_D &=  - \im  {
 7 \over 2 } \epsilon^{ i i_1 \cdots i_4 } \partial_{i_1} \tilde{x}^8
 \partial_{i_2} \tilde{x}^{ 9 } \partial_{i_3} x^{ [\mu_1 }
 \partial_{i_4} x^{ \nu_1] } \partial_i \delta(\sigma-\sigma'), 
\nonumber \\
\left\{ Z^{ [1] \mu_1 }_{\mbox{\tiny NS}}(\sigma) , Z^{ [2] \nu_1 \nu_2
 }_{\mbox{\tiny RR}}(\sigma')  \right\}_D &=  - \im  { 7 \over 6 }
 \epsilon^{ i i_1 \cdots i_4 } \partial_{i_1} Y \partial_{i_2} x^{ [
 \mu_1 } \partial_{i_3} x^{ \nu_1 } \partial_{i_4} x^{ \nu_2] }
 \partial_i \delta(\sigma-\sigma'), 
\nonumber \\
\left\{ Z^{ [1] \mu }_{\mbox{\tiny NS}}(\sigma) , Z^{ [2] \nu
 8 }_{\mbox{\tiny RR}}(\sigma')  \right\}_D &=  
\left\{ Z^{ [1] 8 }_{\mbox{\tiny NS}}(\sigma) , Z^{ [2] \mu
 \nu }_{\mbox{\tiny RR}}(\sigma')  \right\}_D \nonumber\\
&=  - \im  { 7 \over 2
 } \epsilon^{ i i_1 \cdots i_4 } \partial_{i_1}  Y \partial_{i_2} \tilde{x}^{
 8 } \partial_{i_3} x^{ [\mu } \partial_{i_4} x^{ \nu] }
 \partial_i \delta(\sigma-\sigma'), 
\nonumber \\
\left\{ Z^{ [1] \mu }_{\mbox{\tiny NS}}(\sigma) , Z^{ [2] \nu
 9 }_{\mbox{\tiny RR}}(\sigma')  \right\}_D &=  
\left\{ Z^{ [1] 9 }_{\mbox{\tiny NS}}(\sigma) , Z^{ [2] \mu
 \nu }_{\mbox{\tiny RR}}(\sigma')  \right\}_D \nonumber\\
&=  - \im  { 7 \over 2
 } \epsilon^{ i i_1 \cdots i_4 } \partial_{i_1}  Y \partial_{i_2} \tilde{x}^{
 9 } \partial_{i_3} x^{ [\mu } \partial_{i_4} x^{ \nu] }
 \partial_i \delta(\sigma-\sigma'), 
\nonumber \\
\left\{ Z^{ [1] \mu }_{\mbox{\tiny NS}}(\sigma) , Z^{ [2] 8
 9 }_{\mbox{\tiny RR}}(\sigma')  \right\}_D &=  7 \im 
 \epsilon^{ i i_1 \cdots i_4 } \partial_{i_1} Y \partial_{i_2} \tilde{x}^{
 9 } \partial_{i_3} \tilde{x}^{  8 } \partial_{i_4} x^{ \mu }
 \partial_i \delta(\sigma-\sigma'), 
\nonumber \\
\left\{ Z^{ [1] 8 }_{\mbox{\tiny NS}}(\sigma) , Z^{ [2] \nu
 9 }_{\mbox{\tiny RR}}(\sigma')  \right\}_D &=  7 \im 
 \epsilon^{ i i_1 \cdots i_4 } \partial_{i_1}  Y \partial_{i_2} \tilde{x}^{
 8 } \partial_{i_3} \tilde{x}^{ 9 } \partial_{i_4} x^{ \nu }
 \partial_i \delta(\sigma-\sigma'), 
\nonumber \\
\left\{ Z^{ [1] 9 }_{\mbox{\tiny NS}}(\sigma) , Z^{ [2] \nu 8 }_{\mbox{\tiny RR}}(\sigma')  \right\}_D &=   
7 \im \epsilon^{ i i_1 \cdots i_4 } \partial_{i_1}  Y \partial_{i_2} \tilde{x}^{ 9 } \partial_{i_3} \tilde{x}^{ 8 } \partial_{i_4} x^{ \nu } \partial_i \delta(\sigma-\sigma').
\end{align}
Each component is given by \eqref{eq:IIANS5_current}
, also where the indices $\mu,\nu \cdots \ne 8,9$. 
All the background fields in the momenta $p_{\mu} \, (\mu \not= 8,9), p_8, p_9, p_Y$ are replaced by
the T-dualized ones. 

Again, due to the self-duality constraints for $A^{[2]}$ in the
$\mathcal{N} = (2,0)$ tensor multiplet, the algebra is governed by the
Dirac bracket.
We have two distinguished directions $\tilde{x}^8$ and $\tilde{x}^9$ in
the $5^2_2$-brane.
We stress that the complicated structure of the worldvolume theory is
substantially encoded into the form of the momenta $p_{\mu}, p_8, p_9,
p_Y$.

\section{Current algebras in $\mathcal{N} = (1,1)$ theories}
\label{sect:vector}
In this section, we derive the current algebras of type IIB NS5-brane,
the IIA KK5-brane and the IIB $5^2_2$-brane.
The current algebras of these branes are characterized by the $\mathcal{N} = (1,1)$ vector
multiplet 
whose bosonic components are four scalar fields and a vector field.
In order to write down these algebras, 
we first introduce the algebra of the D5-brane in type IIB theory \cite{Kamimura:1997ju}.
We then perform the S-duality transformation to obtain the current of
the type IIB NS5-brane.

The action for a D$p$-brane is given by the sum of the DBI action and 
 the WZ term as
\bea
S&=&S_{\mbox{\tiny DBI}}+S_{\mbox{\tiny WZ}} ~~,~~S_{\mbox{\tiny DBI}}= T_{Dp} \displaystyle\int_M 
d^p\sigma~{\cal L}_{\mbox{\tiny DBI}},
\nn \\
{\cal L}_{\mbox{\tiny DBI}}&=&-e^{-\phi}\sqrt{-h_{F}}~~,~~
h_{F}=\det h_F{}_{\hat{i} \hat{j}},
\nn \\
S_{\mbox{\tiny WZ}}&=&T_{Dp}\displaystyle\int_M 
e^{{\cal F}} P [C^{\mathrm{RR}}],
\nn \\
h_F{}_{\hat{i} \hat{j}}&=&\partial_{\hat{i}} x^{\mu} \partial_{\hat{j}}
x^{\nu} G_{\mu \nu}
+ {\cal F}_{\hat{i} \hat{j}},
\nn \\
F_{\hat{i} \hat{j}} &=& \partial_{\hat{i}} A_{\hat{j}} - \partial_{\hat{j}} A_{\hat{i}} ~~,~~
{\cal F}_{\hat{i} \hat{j}}~=~F_{\hat{i} \hat{j}}+
\partial_{\hat{i}} x^{\mu} \partial_{\hat{j}} x^{\nu} B_{\mu \nu}
~~~.
\eea
Here $T_{Dp}$ is the tension of the D$p$-brane, 
$C^{\mathrm{RR}}$ is the RR polyform potential, $F = dA$ is the field strength of
the worldvolume vector $A = A_{\hat{i}} d \sigma^{\hat{i}}$.
The effective theory is governed by the six-dimensional $\mathcal{N} =
(1,1)$ vector multiplet.

The canonical momenta conjugate to $x^{\mu}$ and $A_i$ 
are defined as 
\bea
p_{\mu}
&=&-
\sqrt{-h_F}\left(\frac{1}{2} h_F{}^{(\hat{i} 0)} G_{\mu \nu} +
\frac{1}{2}h_F{}^{[\hat{i} 0]}
B_{\mu \nu}\right) \partial_{\hat{i}} x^{\nu}
+\displaystyle\frac{\partial {\cal L}_{\mbox{\tiny WZ}}}{\partial
(\partial_0 x^{\mu})}, 
\nn \\
E^i&=&- 
\sqrt{-h_F}\frac{1}{2}h_F{}^{[i0]}+
\displaystyle\frac{\partial {\cal L}_{WZ}}{\partial F_{0i}}
~~~,~~_{i=1,\cdots,p}~~~.
\eea
The Hamiltonian is given by \cite{Kamimura:1997ju}
\bea
{\cal H}&=& p_{\mu} \partial_0 x^{\mu}
+E^i\partial_0 A_i-{\cal L}
\nn \\
&=&-\frac{1}{\sqrt{-h}h^{00}}{\cal H}_\perp
-\frac{h^{0i}}{h^{00}}{\cal H}_i
-A_0~\Phi,
\eea
where we have defined
\bea
		{\cal H}_\perp &=&\displaystyle\frac{1}{2}e^\phi
		\left(
		\tilde{p}_{\mu} G^{\mu \nu}\tilde{p}_{\nu}
		+
		\tilde{E}^ih_{ij}\tilde{E}^j
		+e^{-2\phi} \det h_F{}_{ij}
		\right)~=~0, \nn\\
		{\cal H}_i&=&\tilde{p}_{\mu} \partial_i  x^{\mu} + {\cal F}_{ij}\tilde{E}^j~=~
		{p}_{\mu} \partial_i  x^{\mu} + {F}_{ij} {E}^j~=~
		0, \nn \\
		\Phi&=&\partial_iE^i~=~0,
\eea
with 
the following quantities
\bea
\tilde{p}_{\mu} & = & p_{\mu}
-B_{\mu \nu}E^i\partial_i x^{\nu}
-\displaystyle\frac{\partial {\cal L}_{\mbox{\tiny WZ}}}{\partial (\partial_0 x^{\mu})},
\nn\\
&=&-e^{-\phi}\sqrt{-h_F}\frac{1}{2}h_F{}^{(\hat{i} 0)}G_{\mu
\nu}\partial_{\hat{i}} x^{\nu}~~,
\nn\\
\tilde{E}^i & = & E^i-
\displaystyle\frac{\partial {\cal L}_{\mbox{\tiny WZ}}}{\partial F_{0i}}
=- e^{-\phi}\sqrt{-h_F}\frac{1}{2} h_F{}^{[i0]}~.
\eea
For the D5-brane in type IIB theory, 
${\cal H}_\perp$
is written by the sum of bilinears \cite{Kamimura:1997ju},
\bea
{\cal H}_\perp&=&\frac{1}{2}
Z_M~{\cal M}^{MN}~Z_N,
\eea
where 
\bea
Z_M&=&\left({\renewcommand{\arraystretch}{1.2}
	\begin{array}{c}
		p_{\mu} \\
		Z_{\mbox{\tiny NS}}^{[1]\mu} \\
		Z_{\mbox{\tiny RR}}^{[1]\mu}\\
		Z_{\mbox{\tiny RR}}^{[3] \mu_1 \mu_2 \mu_3}\\
		Z_{\mbox{\tiny RR}}^{[5] \mu_1 \mu_2 \mu_3 \mu_4 \mu_5}
\end{array}}
\right).
\eea
Each component is explicitly given by
\bea
	    Z_{\mbox{\tiny NS}}^{[1]\mu}&=&E^i\partial_ix^{\mu}, \nonumber \\
	    Z_{\mbox{\tiny RR}}^{[1]\mu}&=&\epsilon^{i_1 i_2 i_3
	     i_4 i_5} { F}_{i_1i_2} { F}_{i_3i_4}
	     \partial_{i_5}x^{\mu}, \nonumber \\
	    Z_{\mbox{\tiny RR}}^{[3] \mu_1 \mu_2 \mu_3}&=&
		\epsilon^{i_1 i_2 i_3 i_4 i_5} { F}_{i_1i_2}
		\partial_{i_3}x^{\mu_1}\partial_{i_4}x^{\mu_2}
	    \partial_{i_5}x^{\mu_{3}}, \nonumber \\
	    Z_{\mbox{\tiny RR}}^{[5] \mu_1 \mu_2 \mu_3 \mu_4 \mu_5}&=&
	\epsilon^{i_1 i_2 i_3 i_4 i_5}\partial_{i_1}x^{\mu_1}\partial_{i_2}x^{\mu_2}
	\partial_{i_3}x^{\mu_3}\partial_{i_4}x^{\mu_4} \partial_{i_5}x^{\mu_5}.
\eea
The current algebra of the D5-brane is therefore given by 
\bea
&&
\left\{Z_M(\sigma),Z_N(\sigma')\right\}=
\im \rho_{MN}^i \partial_i\delta(\sigma-\sigma')\nn\\\nn\\
&&\rho_{MN}^i=\left({\renewcommand{\arraystretch}{1.2}
	\begin{array}{cc|ccc}
		0&
		E^i\delta_{\mu}^{\nu}
		&\rho_{13}&\rho_{14}&\rho_{15}\\
		E^i\delta_{\nu}^{\mu}&0&\rho_{23}&\rho_{24}&0\\\hline
		\rho_{13}		&	\rho_{23}&&\\
		\rho_{14}&\rho_{24}&&\\
		\rho_{15}		&0&&&
\end{array}}
\right)\label{IIBD$5$rho}
\eea
where each non-zero component is given by 
\bea
&&~~~~~~~
\rho_{13}=\epsilon^{ii_1\cdots i_{4}}
F{}_{i_1i_{2}} F{}_{i_3 i_{4}} 
\delta_{\mu}^{\nu}\nn\\
&&~~~~~~~
\rho_{14}=3\epsilon^{ii_1\cdots i_{4}}
F_{i_1 i_{2}} \partial_{i_3}x^{[\nu_1}\partial_{i_4}x^{\nu_2}
\delta_\mu^{\nu]}/3!\nn\\
&&~~~~~~~\rho_{15}=5 \epsilon^{ii_1\cdots i_{4}}
\partial_{i_1}x^{[\nu_1}\partial_{i_2}x^{\nu_2}\partial_{i_3}x^{\nu_3} \partial_{i_{4}}x^{\nu_{4}}\delta_{\mu}^{\nu_{5}]}/5!\nn\\
&&~~~~~~~\rho_{23}=4 \epsilon^{ii_1\cdots i_4}
F_{i_1i_2}\partial_{i_3}x^{\mu}\partial_{i_4}x^{\nu}
\nn\\
&&~~~~~~~\rho_{24}=2 \epsilon^{ii_1\cdots i_{4}}
\partial_{i_1}x^{\mu}\partial_{i_{2}}x^{\nu_{1}}
\partial_{i_3}x^{\nu_2}\partial_{i_{4}}x^{\nu_{3}}.
\label{eq:D5_algebra}
\eea

\subsection{IIB NS5-brane}

The effective action of the type IIB NS5-brane is obtained by the
S-duality transformation of the D5-brane action \cite{Eyras:1998hn} and 
it is governed by the $\mathcal{N} = (1,1)$ vector multiplet.
The S-duality transformation rules of the background fields are
\begin{eqnarray}
\begin{aligned}
\tau \ &\xrightarrow[\text{S}]{} \ 
- \frac{1}{\tau}
\, , \ \ \ \ \
C^{[2]} \ \xrightarrow[\text{S}]{} \ B
\, , &\ \ \ \ \
B \ &\xrightarrow[\text{S}]{} \ - C^{[2]}
\, , \ \ \ \ \
g_{\mu \nu} 
\ \xrightarrow[\text{S}]{} \ 
|\tau| \, g_{\mu \nu}
\, , \\
C^{[4]} \ &\xrightarrow[\text{S}]{} \ 
C^{[4]} + C^{[2]} \wedge B
\, , &\ \ \ \ \ 
C^{[6]} \ &\xrightarrow[\text{S}]{} \ 
- B^{[6]} + \frac{1}{2} B \wedge C^{[2]} \wedge C^{[2]}
\, ,
\end{aligned}
\label{eq:S-duality}
\end{eqnarray}
where $\tau = C^{[0]} + \im  \, e^{-\phi}$ is the complex axion-dilaton
field and $B^{[6]}$ is the magnetic dual of $B$.
The algebra of the type IIB NS5-brane is obtained by performing the
S-duality transformation of that of the D5-brane \eqref{eq:D5_algebra}.
The non-zero components of the current algebra in the type IIB NS5-brane
are therefore 
\begin{align}
\left\{ Z_{\mu}(\sigma), Z^{[1]\nu}_{\mbox{\tiny NS}}(\sigma') \right\}&=  \im E^i \delta_{\mu}^{\nu} \partial_i
 \delta(\sigma-\sigma'),
\nonumber \\
\left\{ Z_{\mu}(\sigma), Z^{[1]\nu}_{\mbox{\tiny RR}}(\sigma') \right\}&=\im
 \epsilon^{i i_1 \cdots i_4} 
 F_{i_1i_2} F_{i_3i_4} \delta^{\nu}_{\mu} \partial_i
 \delta(\sigma-\sigma'), 
\nonumber \\
\left\{ Z_{\mu}(\sigma), Z^{[3]\nu_1\nu_2\nu_3}_{\mbox{\tiny RR}}(\sigma')
 \right\}&= { \im \over 2! } \epsilon^{i i_1 \cdots
 i_4} F_{i_1i_2} \partial_{ i_3 } x^{ [\nu_1 }
 \partial_{i_4} x^{\nu_2} \delta_{\mu}^{\nu_3]} \partial_i
 \delta(\sigma-\sigma'), 
\nonumber \\
\left\{ Z_{\mu}(\sigma), Z^{[5]\nu_1 \cdots \nu_5}_{\mbox{\tiny
 RR}}(\sigma') \right\}&={ \im \over 4! } \epsilon^{i
 i_1 \cdots i_4} \partial_{ i_1 } x^{ [\nu_1 } \cdots \partial_{i_4}
 x^{\nu_4} \delta_{\mu}^{\nu_5]} \partial_i \delta(\sigma-\sigma'), 
\nonumber \\
\left\{ Z^{[1]\mu}_{\mbox{\tiny NS}}(\sigma), Z^{[1]\nu}_{\mbox{\tiny
 RR}}(\sigma') \right\}&=4\im \epsilon^{i
 i_1 \cdots i_4} F_{i_1i_2} \partial_{i_3} x^{\mu} \partial_{i_4}
 x^{\nu} \partial_i \delta(\sigma-\sigma'), 
\nonumber \\
\left\{ Z^{[1]\mu}_{\mbox{\tiny NS}}(\sigma), Z^{[3]\nu_1 \nu_2
 \nu_3}_{\mbox{\tiny RR}}(\sigma') \right\}&=2\im
 \epsilon^{i i_1 \cdots i_4}  \partial_{i_1} x^{\mu} \partial_{i_2}
 x^{\nu_1} \partial_{i_3} x^{\nu_2} \partial_{i_4} x^{\nu_3} \partial_i
 \delta(\sigma-\sigma').
\label{eq:IIBNS5_algebra}
\end{align}
Here each component is given by \eqref{eq:D5_algebra} but all the background fields
in the momentum $p_{\mu}$ are replaced by the S-dualized ones.

Compared with the type IIA NS5-brane, the algebra
\eqref{eq:IIBNS5_algebra} is characterized by the Poisson bracket 
$\{\cdot, \cdot\}$.
This apparently comes from the fact that the theory is governed by the
$\mathcal{N} = (1,1)$ vector multiplet.
There is no second class constraint in the effective
theory and we do not care about the restricted phase space.

\subsection{IIA KK5-brane}
The type IIA KK5-brane is obtained by the T-duality transformation along
a transverse direction to the IIB NS5-brane.
The transformations of the background fields are given by the Buscher
rule \eqref{eq:Buscher_rule}.
Given the worldvolume action of the type IIB NS5-brane, we obtain  
the worldvolume effective action of the type IIA KK5-brane
\cite{Eyras:1998hn}. 
We also recombine the type IIB currents $Z_M$ into those in the type IIA.
	The left hand sides are the IIB RR currents and the right hand sides are the IIA RR currents:
	\begin{displaymath}
		Z^{[1]}_{\mbox {\tiny RR}}
		\left\{
		\begin{array}{l}
		Z^{[1] \mu}_{\mbox {\tiny RR}} = Z^{[2]\mu9}_{\mbox {\tiny RR}}\\	
                Z^{[1] 9}_{\mbox {\tiny RR}} = Z^{[0]}_{\mbox {\tiny RR}}\\
		\end{array}
		\right.,~~~
		Z^{[3]}_{\mbox{ \tiny RR }}
		\left\{
		\begin{array}{l}
			Z^{[3] \mu\nu\rho}_{\mbox{ \tiny RR }} = Z^{ [4] \mu\nu\rho9}_{ \mbox{ \tiny RR } }\\
			Z^{[3] \mu\nu9}_{\mbox{ \tiny RR }} = Z^{[2]\mu\nu}_{\mbox{ \tiny RR }}
		\end{array}
                \right.,~~~
		Z^{ [5+] \mu_1 \cdots \mu_4 9}_{ \mbox{ \tiny RR } }
		=	Z^{[4] \mu_1 \cdots \mu_4 }_{\mbox{ \tiny RR }}
		.
	\end{displaymath}
We find that the non-zero components of the IIA KK5-brane algebra are
given by
\begin{align}
	\left\{ Z_{\mu}(\sigma), Z^{[1]\nu}_{\mbox{\tiny NS}}(\sigma') \right\}&= \im E^i \delta_{\mu}^{\nu} \partial_i
	\delta(\sigma-\sigma'),
	\nonumber \\
	\left\{ Z_{9}(\sigma),  Z^{[1]9}_{\mbox{\tiny NS}}(\sigma')
	\right\}&=\im  E^i \partial_i
	\delta(\sigma-\sigma'),
	\nonumber \\
	\left\{ Z_{\mu}(\sigma), Z^{[2]\nu9}_{\mbox{\tiny RR}}(\sigma') \right\}&=\im
	\epsilon^{i i_1 \cdots i_4} 
	F_{i_1i_2} F_{i_3i_4} \delta^{\nu}_{\mu} \partial_i
	\delta(\sigma-\sigma'), 
	\nonumber \\
	\left\{ Z_{9}(\sigma), Z^{[0]}_{\mbox{\tiny RR}}(\sigma')
	\right\}&=  \im\epsilon^{i i_1 \cdots i_4}  F_{i_1i_2} F_{i_3i_4} \partial_i \delta(\sigma-\sigma'), 
	\nonumber \\
	\left\{ Z_{\mu}(\sigma), Z^{[4]\nu_1\nu_2\nu_39}_{\mbox{\tiny RR}}(\sigma')
	\right\}&= { \im \over 2! } \epsilon^{i i_1 \cdots
		i_4} F_{i_1i_2} \partial_{ i_3 } x^{ [\nu_1 }
	\partial_{i_4} x^{\nu_2} \delta_{\mu}^{\nu_3]} \partial_i
	\delta(\sigma-\sigma'), 
	\nonumber \\
	\left\{ Z_{\mu}(\sigma), Z^{[2]\nu_1\nu_2}_{\mbox{\tiny
			RR}}(\sigma') \right\}&= \im \epsilon^{i i_1 \cdots
		i_4} F_{i_1i_2} \partial_{ i_3 } \tilde{x}^{ 9 }
	\partial_{i_4} x^{[\nu_1} \delta_{\mu}^{\nu_2]} \partial_i
	\delta(\sigma-\sigma'), 
	\nonumber \\
	\left\{ Z_{9}(\sigma), Z^{[2]\nu_1\nu_2 }_{\mbox{\tiny
			RR}}(\sigma') \right\}&= { \im \over 2! } \epsilon^{i
		i_1 \cdots i_4} F_{i_1i_2} \partial_{ i_3 } x^{ [\nu_1 }
	\partial_{i_4} x^{\nu_2]} \partial_i \delta(\sigma-\sigma'), 
	\nonumber \\
	\left\{ Z_{\mu}(\sigma), Z^{[4]\nu_1 \cdots \nu_4}_{\mbox{\tiny
			RR}}(\sigma') \right\}&= { \im \over 3! } \epsilon^{i
		i_1 \cdots i_4} \partial_{ i_1 } \tilde{x}^{ 9 } \partial_{i_2} x^{
		[\nu_1 }\partial_{i_3} x^{\nu_2}  \partial_{i_4} x^{\nu_3}
	\delta_{\mu}^{\nu_4]} \partial_i \delta(\sigma-\sigma'), 
	\nonumber \\
	\left\{ Z_{9}(\sigma), Z^{[4]\nu_1 \cdots \nu_4
		}_{\mbox{\tiny RR}}(\sigma') \right\}&= { \im \over 4!
	} \epsilon^{i i_1 \cdots i_4} \partial_{ i_1 } x^{
		[\nu_1 } \cdots \partial_{i_4} x^{\nu_4]} \partial_i
	\delta(\sigma-\sigma'), 
	\nonumber \\
	\left\{ Z^{[1]\mu}_{\mbox{\tiny NS}}(\sigma), Z^{[2]\nu9}_{\mbox{\tiny
			RR}}(\sigma') \right\}&=4\im  \epsilon^{i
		i_1 \cdots i_4} F_{i_1i_2} \partial_{i_3} x^{\mu} \partial_{i_4}
	x^{\nu} \partial_i \delta(\sigma-\sigma'), 
	\nonumber \\
	\left\{ Z^{[1]\mu}_{\mbox{\tiny NS}}(\sigma), Z^{[0]}_{\mbox{\tiny
			RR}}(\sigma') \right\}&=4\im  \epsilon^{i
		i_1 \cdots i_4} F_{i_1i_2} \partial_{i_3} x^{\mu} \partial_{i_4}
	\tilde{x}^9 \partial_i \delta(\sigma-\sigma'), 
	\nonumber \\
	\left\{ Z^{[1]9}_{\mbox{\tiny NS}}(\sigma), Z^{[2]\nu9}_{\mbox{\tiny
			RR}}(\sigma') \right\}&=4\im  \epsilon^{i
		i_1 \cdots i_4} F_{i_1i_2} \partial_{i_3} \tilde{x}^9 \partial_{i_4}
	x^{\nu} \partial_i \delta(\sigma-\sigma'), 
	\nonumber \\
	\left\{ Z^{[1]\mu}_{\mbox{\tiny NS}}(\sigma), Z^{[4]\nu_1 \nu_2
		\nu_39}_{\mbox{\tiny RR}}(\sigma') \right\}&=2\im
	\epsilon^{i i_1 \cdots i_4}  \partial_{i_1} x^{\mu} \partial_{i_2}
	x^{\nu_1} \partial_{i_3} x^{\nu_2} \partial_{i_4} x^{\nu_3} \partial_i
	\delta(\sigma-\sigma'), 
	\nonumber \\
	\left\{ Z^{[1]9}_{\mbox{\tiny NS}}(\sigma), Z^{[4]\nu_1 \nu_2
		\nu_39}_{\mbox{\tiny RR}}(\sigma') \right\}&=2\im
	\epsilon^{i i_1 \cdots i_4}  \partial_{i_1} \tilde{x}^9
	\partial_{i_2} x^{\nu_1} \partial_{i_3} x^{\nu_2} \partial_{i_4}
	x^{\nu_3} \partial_i \delta(\sigma-\sigma'), 
	\nonumber \\
	\left\{ Z^{[1]\mu}_{\mbox{\tiny NS}}(\sigma),
	Z^{[2]\nu_1 \nu_2}_{\mbox{\tiny RR}}(\sigma')
	\right\}&=2\im \epsilon^{i i_1 \cdots i_4}
	\partial_{i_1} x^{\mu} \partial_{i_2} x^{\nu_1} \partial_{i_3}
	x^{\nu_2} \partial_{i_4} \tilde{x}^9 \partial_i
	\delta(\sigma-\sigma'),
	\label{eq:IIAKK5_algebra}\end{align}
where each component is defined by \eqref{eq:D5_algebra} but all the background fields
in the momenta $p_{\mu}, p_9 \ (\mu \not= 9)$ are replaced by the T-dualized ones.

\subsection{IIB $5^2_2$-brane}
A further T-duality transformation to another transverse isometric direction to
the type IIB KK5-brane results in the exotic $5^2_2$-brane in type IIB theory.
The worldvolume effective action of the type IIB $5^2_2$-brane has been
obtained in \cite{Chatzistavrakidis:2013jqa,Kimura:2014upa}.
The non-zero components of the type IIB $5^2_2$-brane algebra are 
\begin{align}
\left\{ Z_{\mu}(\sigma), Z^{[1]\nu}_{\mbox{\tiny NS}}(\sigma') \right\}&=\im  E^i \delta_{\mu}^{\nu} \partial_i
 \delta(\sigma-\sigma'),
\nonumber \\
\left\{ Z_{9}(\sigma),  Z^{[1]9}_{\mbox{\tiny NS}}(\sigma')
 \right\}&= 
\left\{ Z_{8}(\sigma), Z^{[1]8}_{\mbox{\tiny NS}}(\sigma')
 \right\}=  \im E^i \partial_i
 \delta(\sigma-\sigma'),
\nonumber \\
\left\{ Z_{\mu}(\sigma), Z^{[1]\nu}_{\mbox{\tiny RR}}(\sigma') \right\}&=\im
  \epsilon^{i i_1 \cdots i_4} 
 F_{i_1i_2} F_{i_3i_4} \delta^{\nu}_{\mu} \partial_i
 \delta(\sigma-\sigma'), 
\nonumber \\
\left\{ Z_{9}(\sigma), Z^{[1]9}_{\mbox{\tiny RR}}(\sigma')
 \right\}&= 
\left\{ Z_{8}(\sigma), Z^{[1]8}_{\mbox{\tiny RR}}(\sigma')
 \right\}= \im \epsilon^{i i_1 \cdots i_4}  F_{i_1i_2} F_{i_3i_4} \partial_i \delta(\sigma-\sigma'), 
\nonumber \\
\left\{ Z_{\mu}(\sigma), Z^{[3]\nu_1\nu_2\nu_3}_{\mbox{\tiny RR}}(\sigma')
 \right\}&= { \im \over 2! } \epsilon^{i i_1 \cdots
 i_4} F_{i_1i_2} \partial_{ i_3 } x^{ [\nu_1 }
 \partial_{i_4} x^{\nu_2} \delta_{\mu}^{\nu_3]} \partial_i
 \delta(\sigma-\sigma'), 
\nonumber \\
\left\{ Z_{\mu}(\sigma), Z^{[3]\nu_1\nu_2 9}_{\mbox{\tiny
 RR}}(\sigma') \right\}&=\im  \epsilon^{i i_1 \cdots
 i_4} F_{i_1i_2} \partial_{ i_3 } \tilde{x}^{ 9 }
 \partial_{i_4} x^{[\nu_1} \delta_{\mu}^{\nu_2]} \partial_i
 \delta(\sigma-\sigma'), 
\nonumber \\
\left\{ Z_{\mu}(\sigma), Z^{[3]\nu_1\nu_2 9}_{\mbox{\tiny
 RR}}(\sigma') \right\}&= \im \epsilon^{i i_1 \cdots
 i_4} F_{i_1i_2} \partial_{ i_3 } \tilde{x}^{ 8 }
 \partial_{i_4} x^{[\nu_1} \delta_{\mu}^{\nu_2]} \partial_i
 \delta(\sigma-\sigma'), 
\nonumber \\
\left\{ Z_{9}(\sigma), Z^{[3]\nu_1\nu_2 9}_{\mbox{\tiny
 RR}}(\sigma') \right\}&= 
\left\{ Z_{8}(\sigma), Z^{[3]\nu_1\nu_2 8}_{\mbox{\tiny
 RR}}(\sigma') \right\} \nonumber\\
&= { \im \over 2! } \epsilon^{i
 i_1 \cdots i_4} F_{i_1i_2} \partial_{ i_3 } x^{ [\nu_1 }
 \partial_{i_4} x^{\nu_2]} \partial_i \delta(\sigma-\sigma'), 
\nonumber \\
\left\{ Z_{\mu}(\sigma), Z^{[3]\nu_1 8 9}_{\mbox{\tiny
 RR}}(\sigma') \right\}&= \im \epsilon^{i i_1 \cdots
 i_4} F_{i_1i_2} \partial_{ i_3 } \tilde{x}^{ 8 }
 \partial_{i_4} \tilde{x}^9 \delta_{\mu}^{\nu_1} \partial_i
 \delta(\sigma-\sigma'), 
\nonumber \\
\left\{ Z_{9}(\sigma), Z^{[3]\nu_1 8 9}_{\mbox{\tiny
 RR}}(\sigma') \right\}&= \im \epsilon^{i i_1 \cdots
 i_4} F_{i_1i_2} \partial_{ i_3 } x^{ \nu_1 }
 \partial_{i_4} \tilde{x}^8 \partial_i \delta(\sigma-\sigma'), 
\nonumber \\
\left\{ Z_{8}(\sigma), Z^{[3]\nu_1 9 8}_{\mbox{\tiny RR}}(\sigma')
 \right\}&=\im \epsilon^{i i_1 \cdots i_4}
 F_{i_1i_2} \partial_{ i_3 } x^{ \nu_1 } \partial_{i_4} \tilde{x}^9
 \partial_i \delta(\sigma-\sigma'), 
\nonumber \\
\left\{ Z_{\mu}(\sigma), Z^{[5]\nu_1 \cdots \nu_5}_{\mbox{\tiny
 RR}}(\sigma') \right\}&= { \im \over 4! } \epsilon^{i
 i_1 \cdots i_4} \partial_{ i_1 } x^{ [\nu_1 } \cdots \partial_{i_4}
 x^{\nu_4} \delta_{\mu}^{\nu_5]} \partial_i \delta(\sigma-\sigma'), 
\nonumber \\
\left\{ Z_{\mu}(\sigma), Z^{[5]\nu_1 \cdots \nu_4 9}_{\mbox{\tiny
 RR}}(\sigma') \right\}&= { \im \over 3! } \epsilon^{i
 i_1 \cdots i_4} \partial_{ i_1 } \tilde{x}^{ 9 } \partial_{i_2} x^{
 [\nu_1 }\partial_{i_3} x^{\nu_2}  \partial_{i_4} x^{\nu_3}
 \delta_{\mu}^{\nu_4]} \partial_i \delta(\sigma-\sigma'), 
\nonumber \\
\left\{ Z_{\mu}(\sigma), Z^{[5]\nu_1 \cdots \nu_4 8}_{\mbox{\tiny
 RR}}(\sigma') \right\}&= { \im \over 3! } \epsilon^{i
 i_1 \cdots i_4} \partial_{ i_1 } \tilde{x}^{ 8 } \partial_{i_2} x^{
 [\nu_1 }\partial_{i_3} x^{\nu_2}  \partial_{i_4} x^{\nu_3}
 \delta_{\mu}^{\nu_4]} \partial_i \delta(\sigma-\sigma'), 
\nonumber \\
\left\{ Z_{9}(\sigma), Z^{[5]\nu_1 \cdots \nu_4
 9}_{\mbox{\tiny RR}}(\sigma') \right\}&= 
\left\{ Z_{8}(\sigma), Z^{[5]\nu_1 \cdots \nu_4
 8}_{\mbox{\tiny RR}}(\sigma') \right\}= { \im \over 4!
 } \epsilon^{i i_1 \cdots i_4} \partial_{ i_1 } x^{
 [\nu_1 } \cdots \partial_{i_4} x^{\nu_4]} \partial_i
 \delta(\sigma-\sigma'), 
\nonumber \\
\left\{ Z_{\mu}(\sigma), Z^{[5]\nu_1 \nu_2 \nu_3 8
 9}_{\mbox{\tiny RR}}(\sigma') \right\}&= { \im \over 2!
 } \epsilon^{i i_1 \cdots i_4} \partial_{ i_1 } \tilde{x}^{
 8 } \partial_{i_2} \tilde{x}^{ 9 }\partial_{i_3} x^{[\nu_1}
 \partial_{i_4} x^{\nu_2} \delta_{\mu}^{\nu_3]} \partial_i
 \delta(\sigma-\sigma'), 
\nonumber \\
\left\{ Z_{9}(\sigma), Z^{[5]\nu_1 \nu_2 \nu_3 8
 9}_{\mbox{\tiny RR}}(\sigma') \right\}&= { \im \over 3!
 } \epsilon^{i i_1 \cdots i_4} \partial_{ i_1 } x^{
 [\nu_1 } \partial_{i_2} x^{\nu_2} \partial_{i_3} x^{\nu_3]}
 \partial_{i_4} \tilde{x}^8 \partial_i \delta(\sigma-\sigma'), 
\nonumber \\
\left\{ Z_{8}(\sigma), Z^{[5]\nu_1 \nu_2 \nu_3 9
 8}_{\mbox{\tiny RR}}(\sigma') \right\}&= { \im \over 3!
 } \epsilon^{i i_1 \cdots i_4} \partial_{ i_1 } x^{
 [\nu_1 } \partial_{i_2} x^{\nu_2} \partial_{i_3} x^{\nu_3]}
 \partial_{i_4} \tilde{x}^9 \partial_i \delta(\sigma-\sigma'), 
\nonumber\\
\left\{ Z^{[1]\mu}_{\mbox{\tiny NS}}(\sigma), Z^{[1]\nu}_{\mbox{\tiny
 RR}}(\sigma') \right\}&=4 \im \epsilon^{i
 i_1 \cdots i_4} F_{i_1i_2} \partial_{i_3} x^{\mu} \partial_{i_4}
 x^{\nu} \partial_i \delta(\sigma-\sigma'), 
\nonumber \\
\left\{ Z^{[1]\mu}_{\mbox{\tiny NS}}(\sigma), Z^{[1]9}_{\mbox{\tiny
 RR}}(\sigma') \right\}&=4 \im \epsilon^{i
 i_1 \cdots i_4} F_{i_1i_2} \partial_{i_3} x^{\mu} \partial_{i_4}
 \tilde{x}^9 \partial_i \delta(\sigma-\sigma'), 
\nonumber \\
\left\{ Z^{[1]9}_{\mbox{\tiny NS}}(\sigma), Z^{[1]\nu}_{\mbox{\tiny
 RR}}(\sigma') \right\}&=4 \im \epsilon^{i
 i_1 \cdots i_4} F_{i_1i_2} \partial_{i_3} \tilde{x}^9 \partial_{i_4}
 x^{\nu} \partial_i \delta(\sigma-\sigma'), 
\nonumber \\
\left\{ Z^{[1]\mu}_{\mbox{\tiny NS}}(\sigma), Z^{[1]8}_{\mbox{\tiny
 RR}}(\sigma') \right\}&=4 \im \epsilon^{i
 i_1 \cdots i_4} F_{i_1i_2} \partial_{i_3} x^{\mu} \partial_{i_4}
 \tilde{x}^8 \partial_i \delta(\sigma-\sigma'), 
\nonumber \\
\left\{ Z^{[1]8}_{\mbox{\tiny NS}}(\sigma), Z^{[1]\nu}_{\mbox{\tiny
 RR}}(\sigma') \right\}&=4 \im \epsilon^{i
 i_1 \cdots i_4} F_{i_1i_2} \partial_{i_3} \tilde{x}^8 \partial_{i_4}
 x^{\nu} \partial_i \delta(\sigma-\sigma'), 
\nonumber \\
\left\{ Z^{[1]9}_{\mbox{\tiny NS}}(\sigma),
 Z^{[1]8}_{\mbox{\tiny RR}}(\sigma') \right\}&=4\im
  \epsilon^{i i_1 \cdots i_4} F_{i_1i_2} \partial_{i_3}
 \tilde{x}^9 \partial_{i_4} \tilde{x}^8 \partial_i
 \delta(\sigma-\sigma'), 
\nonumber \\
\left\{ Z^{[1]8}_{\mbox{\tiny NS}}(\sigma),
 Z^{[1]9}_{\mbox{\tiny RR}}(\sigma') \right\}&=4\im
  \epsilon^{i i_1 \cdots i_4} F_{i_1i_2} \partial_{i_3}
 \tilde{x}^8 \partial_{i_4} \tilde{x}^9 \partial_i
 \delta(\sigma-\sigma'), 
\nonumber \\
\left\{ Z^{[1]\mu}_{\mbox{\tiny NS}}(\sigma), Z^{[3]\nu_1 \nu_2
 \nu_3}_{\mbox{\tiny RR}}(\sigma') \right\}&=2\im
 \epsilon^{i i_1 \cdots i_4}  \partial_{i_1} x^{\mu} \partial_{i_2}
 x^{\nu_1} \partial_{i_3} x^{\nu_2} \partial_{i_4} x^{\nu_3} \partial_i
 \delta(\sigma-\sigma'), 
\nonumber \\
\left\{ Z^{[1]9}_{\mbox{\tiny NS}}(\sigma), Z^{[3]\nu_1 \nu_2
 \nu_3}_{\mbox{\tiny RR}}(\sigma') \right\}&=2\im
 \epsilon^{i i_1 \cdots i_4}  \partial_{i_1} \tilde{x}^9
 \partial_{i_2} x^{\nu_1} \partial_{i_3} x^{\nu_2} \partial_{i_4}
 x^{\nu_3} \partial_i \delta(\sigma-\sigma'), 
\nonumber \\
\left\{ Z^{[1]\mu}_{\mbox{\tiny NS}}(\sigma), Z^{[3]\nu_1 \nu_2
 9}_{\mbox{\tiny RR}}(\sigma') \right\}&=2\im
 \epsilon^{i i_1 \cdots i_4}  \partial_{i_1} x^{\mu} \partial_{i_2}
 x^{\nu_1} \partial_{i_3} x^{\nu_2} \partial_{i_4} \tilde{x}^9
 \partial_i \delta(\sigma-\sigma'), 
\nonumber \\
\left\{ Z^{[1]8}_{\mbox{\tiny NS}}(\sigma), Z^{[3]\nu_1 \nu_2
 \nu_3}_{\mbox{\tiny RR}}(\sigma') \right\}&=2\im
 \epsilon^{i i_1 \cdots i_4}  \partial_{i_1} \tilde{x}^8
 \partial_{i_2} x^{\nu_1} \partial_{i_3} x^{\nu_2} \partial_{i_4}
 x^{\nu_3} \partial_i \delta(\sigma-\sigma'), 
\nonumber \\
\left\{ Z^{[1]\mu}_{\mbox{\tiny NS}}(\sigma), Z^{[3]\nu_1 \nu_2
 8}_{\mbox{\tiny RR}}(\sigma') \right\}&=2\im
 \epsilon^{i i_1 \cdots i_4}  \partial_{i_1} x^{\mu} \partial_{i_2}
 x^{\nu_1} \partial_{i_3} x^{\nu_2} \partial_{i_4} \tilde{x}^8
 \partial_i \delta(\sigma-\sigma'),
\nonumber \\
\left\{ Z^{[1]\mu}_{\mbox{\tiny NS}}(\sigma), Z^{[3]\nu_1 8
 9}_{\mbox{\tiny RR}}(\sigma') \right\}&=2\im
 \epsilon^{i i_1 \cdots i_4}  \partial_{i_1} x^{\mu} \partial_{i_2}
 x^{\nu_1} \partial_{i_3} \tilde{x}^8 \partial_{i_4} \tilde{x}^9
 \partial_i \delta(\sigma-\sigma'), 
\nonumber \\
\left\{ Z^{[1]9}_{\mbox{\tiny NS}}(\sigma), Z^{[3]\nu_1 \nu_2
 8}_{\mbox{\tiny RR}}(\sigma') \right\}&=2\im
 \epsilon^{i i_1 \cdots i_4}  \partial_{i_1} \tilde{x}^9
 \partial_{i_2} x^{\nu_1} \partial_{i_3} x^{\nu_2} \partial_{i_4}
 \tilde{x}^8 \partial_i \delta(\sigma-\sigma'), 
\nonumber \\
\left\{ Z^{8[1]}_{\mbox{\tiny NS}}(\sigma), Z^{[3]\nu_1 \nu_2
 9}_{\mbox{\tiny RR}}(\sigma') \right\}&=2\im
 \epsilon^{i i_1 \cdots i_4}  \partial_{i_1} \tilde{x}^8
 \partial_{i_2} x^{\nu_1} \partial_{i_3} x^{\nu_2} \partial_{i_4}
 \tilde{x}^9 \partial_i \delta(\sigma-\sigma'),
\end{align}
where each component is given by \eqref{eq:D5_algebra} but all the background fields
in the momenta $p_{\mu}, p_8, p_9 \ (\mu \not= 8,9)$ are replaced by the T-dualized ones.

We have now written down the current algebras of the five-branes in type
II string theories.
A comment on the furture T-duality transformations is in order.
Since there are two yet untouched transverse directions to the five-branes, 
there ramain two further T-dualized five-branes.
They are denoted as $5^3_2$- and $5^4_2$-branes and have distinct nature
compared with the $5^b_2$-branes $(b=0,1,2)$ discussed above.
We will address these issues in section \ref{sect:conclusion}.

In the following, we show that 
the worldvolume currents of five-branes
give rise to the notion of
the extended (doubled or exceptional) spaces in DFT and EFT and they 
are directly tied with the supersymmetry algebras.

\section{Extended spaces from current algebras } \label{sect:SUSY_algebras}

In this section, after the review of the current algebra approach \cite{Siegel:1993xq, Siegel:1993th, Siegel:1993bj, Hatsuda:2014qqa} to the double field theory, 
we extend this approach to the exceptional field theory by using brane current algebras. 
We focus on brane currents which correspond to brane charges in superalgebras, since a superalgebra includes not only perturbative states but also non-perturbative states representing S-duality.
Supersymmetric branes in ten-dimensional ${\cal N}=2$ theories are classified in 
\cite{Townsend:1995gp , Hull:1997kt} by
 the superalgebras of 32 supercharges $Q_\alpha$ 
\bea
\{Q_\alpha,Q_\beta\}=Z_{\alpha\beta}=Z_{M}\Gamma^M{}_{\alpha\beta}~~~.
\eea
The index $M$ is decomposed by the form  and
 $(\Gamma^M)_{\alpha\beta}$ is anti-symmetric gamma matrices depending on theories.
The bosonic subalgebra $Z_{M}$ includes not only  the momentum and the winding mode but also 
brane charges, such as KK5-brane, NS5-brane, D5-brane.
Exotic branes are variations of them obtained by T-dualities.  
We consider extended spaces spanned by $Z_M$.
Coordinates are $X^{M}$ corresponding to  $Z_M=\frac{1}{\im}\frac{\partial}{\partial X^M}$. 
We derive section conditions and generalized Courant brackets
from brane current algebras where the Courant bracket from the M5-brane in the five-dimensional torus space is the one for $E_{5(5)}$ \cite{Hatsuda:2013dya}. 
The bracket gives  generalized coordinate transformations
in the extended space. 
An analysis on the conserved current in the doubled formalism is found
in \cite{Blair:2015eba}.

\subsection{Doubled space from the string current algebra}

The string current algebra generated by the current $Z_M(\sigma)=(p_\mu,\partial_\sigma x^\mu)$ is given by
\bea
\{Z_M(\sigma),Z_N(\sigma')\}=\im
\eta_{MN}\partial_\sigma\delta(\sigma-\sigma')~.
\label{eq:canonical_bracket}
\eea
Virasoro constraints for a string in curved backgrounds are written in bilinear of bosonic  currents $Z_M$ and the vielbein $E_A{}^M(X^N)$ as
\bea
\left\{
\begin{array}{ccl}
{\cal H}_\perp&=&
\frac{1}{2}Z_M{\cal M}^{MN}Z_N=
\frac{1}{2}Z_ME_A{}^M\hat{\eta}^{AB}E_B{}^NZ_N\\
{\cal H}_\sigma&=&\frac{1}{2}Z_M{\eta}^{MN}Z_N
\end{array}
\right.
\eea
where $\hat{\eta}^{AB}$ is 
the doubled D-dimensional Minkowski metric and $\eta^{MN}$ is the $O$(D,D) invariant metric.
The vielbein satisfies the orthogonal condition $E_A{}^ME_B{}^N\eta_{MN}=\eta_{AB}$.
The index $M$ is raised and lowered by $\eta_{MN}$ and $\eta^{MN}$.
The vielbein is a coset element of $O$(D,D)/$SO$(D$-$1,1)$^2$ which is parametrized as
\bea
E_A{}^M=
\left(\begin{array}{cc}	e_a{}^\nu&0\\0&e_\nu{}^a\end{array}\right)
\left(\begin{array}{cc}	\delta_\nu{}^\mu&B_{\nu\mu}\\0&\delta_\mu{}^\nu\end{array}\right).
\eea
T-duality is $O$(D,D) rotation which acts linearly in $Z_M$ basis.

For a doubled space function $\Phi (X)$ the $\sigma$ component of the
Virasoro operator defines the $\sigma$ derivative as
\bea
 \partial_\sigma \Phi(X^M)=\im \{\int d\sigma' {\cal H}_\sigma(\sigma'),\Phi(X(\sigma))\}~.
\eea
As is well known, the $\sigma$-translational invariance of closed
strings results in the level matching condition for physical states.
The section condition on the doubled space functions $\Phi(X)$ and $\Psi(X)$ are 
imposed  to guarantees that
the physical quantities are inert under the $\sigma$-diffeomorphism
\bea
\partial_M\eta^{MN}\partial_N=0 \Rightarrow 
\partial_M\eta^{MN}\partial_N \Phi(X)=\partial_M\Phi(X)\eta^{MN}\partial_N\Psi(X)=0~.
\eea 
This is nothing but the weak and the strong constraints in DFT
\cite{Siegel:1993th, Siegel:1993bj, Hull:2009mi}.
Introducing components of the doubled coordinate $X^M=(x^\mu,~\tilde{x}_\mu)$
the section condition is written as
\bea
\partial_\mu~ \frac{\partial}{\partial \tilde{x}_{\mu}}=0~.
\label{eq:section_condition_DFT}
\eea

For two vectors in the doubled space $\Lambda_i{}^M$ $i=1,2$
the canonical bracket 
\eqref{eq:canonical_bracket}
between $\Lambda_i{}^MZ_M$ is calculated as follows
\bea
&&\{\Lambda_1^MZ_M(\sigma),\Lambda_2^NZ_N(\sigma')\}\nn\\
\quad\quad\quad\quad  &=&
\frac{1}{\im }\left(
\Lambda_{[1}{}^N\partial_N\Lambda_{2]}{}^M
-\frac{1-K}{2}\Lambda_1{}^N\partial^M\Lambda_{2;N}
+\frac{1+K}{2}(\partial^M\Lambda_1{}^N)\Lambda_{2;N}
\right)Z_M\delta(\sigma-\sigma')\nn\\
\quad\quad\quad\quad &&+\im\left(
\frac{1-K}{2}\Lambda_1{}^N\Lambda_{2;N}(\sigma)
+\frac{1+K}{2}\Lambda_1{}^N\Lambda_{2;N}(\sigma')
\right)\partial_\sigma\delta(\sigma-\sigma')
\label{eq:canonical_bracket2}
\eea
where we have introduced an arbitrary real number $K$.
It is known that the structures of the Courant and the Dorfman brackets
are extracted from the regular part of the current algebra
\eqref{eq:canonical_bracket2} 
\cite{Hatsuda:2012uk}. Namely,
the Courant bracket of two vectors $\hat{\Lambda}_i=\Lambda_i{}^MZ_M$ is
obtained from the 
second line of the current algebra 
\eqref{eq:canonical_bracket2} 
with $K=0$ as
\bea
[\hat{\Lambda}_1,\hat{\Lambda}_2]_C=\frac{1}{\im}\hat{\Lambda}_{12}~,~
\Lambda_{12}{}^M=\Lambda_{[1}{}^N\partial_N\Lambda_{2]}{}^M
-\frac{1}{2}\Lambda_{[1}\partial^M\Lambda_{2];N}~.
\eea
The choise $K=0$ guarantees the anti-symmetric nature of the Courant bracket.
Similarly, 
the Dorfman bracket is given with $K=1$ as
\bea
[\hat{\Lambda}_1,\hat{\Lambda}_2]_D=\frac{1}{\im}\hat{\Lambda}_{12}~,~
\Lambda_{12}{}^M=\Lambda_{[1}{}^N\partial_N\Lambda_{2]}{}^M
+(\partial^M\Lambda_{1}{}^N)\Lambda_{2;N}~.
\eea

The generalized coordinate transformation is obtained by these brackets with the gauge parameter.
Since the D-dimensional vector component of the vielbein field  $E_a{}^M=(e_a{}^\mu,~
e_a{}^\nu B_{\nu\mu})$ is enough to determine the gauge transformation of the vielbein,
we calculate the following canonical commutator between the gauge parameter vector $\Lambda^M=(\lambda^\mu,~\lambda_\mu)$ and $E_a{}^M$ as
\bea
\delta_\lambda E_a{}^MZ_M(\sigma)&=&\im \{\int d\sigma' \lambda^NZ_N(\sigma'),
E_a{}^MZ_M(\sigma)\}\nn\\
\delta_\lambda E_a{}^M&=&\lambda^N\partial_NE_a{}^M
-E_a{}^N\partial_N\lambda^M
+ E_a{}^N\partial_L\lambda^K\eta_{NK}\eta^{ML} ~.
\eea
In component we have
\bea
\left\{\begin{array}{ccl}
\delta_\lambda e_a{}^\mu&=&{\cal L}_\lambda e_a{}^\mu
+(\lambda_\nu\partial^\nu e_a{}^\mu+e_a{}^\nu\partial^\mu\lambda_\nu)
+e_a{}^\rho B_{\rho\nu}\partial^{[\mu}\lambda^{\nu]} \\
\delta_\lambda B_{\rho\nu}&=&{\cal L}_\lambda B_{\rho\nu}
-\partial_{[\rho}\lambda_{\nu]}
+\left(\lambda_\nu\partial^\nu B_{\rho\mu}
-B_{\rho\nu}\partial^\nu\lambda_\mu
-B_{\nu\mu}\partial^\nu \lambda_\rho \right)
+B_{\rho\nu}B_{\sigma\mu}\partial^{[\nu}\lambda^{\sigma]}
\end{array}\right.\label{dgdB}
\eea
where ${\cal L_\lambda}$ is the usual Lie derivative.
After solving the section conditions by setting 
$
\frac{\partial}{\partial \tilde{x}_\mu}=0$, 
the generalized coordinate transformations \bref{dgdB} 
reduces to the usual gauge transformations.

\subsection{Supersymmetry algebras in ten dimensions}

We begin by the eleven-dimensional ${\cal N}=1$ superalgebra  which is  given by 
\bea
\{Q_\alpha,Q_\beta\}&=&
P_m(C\gamma^m)_{\alpha\beta}
+\frac{1}{2!}Z^{[2]mn}(C\gamma_{mn})_{\alpha\beta}\nn\\&&
+\frac{1}{5!}Z^{[5]m_1m_2m_3m_4m_5}(C\gamma_{m_1m_2m_3m_4m_5})_{\alpha\beta},
\eea
where 
$C$ is the charge conjugation matrix.
The number of  bosonic charges, $P_m$, $Z^{[2]}$ and $Z^{[5]}$,
is $11+55+462=32\times 33/2=528$.
$Z^{[2]}$ gives a M2-brane, while $Z^{[5]}$ gives a M5-brane
and a 6-brane.

In $D=10$ the IIA theory has two 10-dimensional Majorana-Weyl supercharges 
which have the opposite chirality, 
$Q_{A\alpha}$ with $A=1,2$ and $\alpha=1,\cdots,16$.
The IIA superalgebra  is given by
\bea
\{Q_{A\alpha},Q_{B\beta}\}&=&
P_{\mu} \delta_{AB}(C\gamma^{\mu})_{\alpha\beta}
+\frac{1}{5!}Z^{[5(+)] \mu_1 \mu_2 \mu_3 \mu_4 \mu_5}_{\mbox{\tiny KK}}\delta_{AB}
(C\gamma_{\mu_1 \mu_2 \mu_3 \mu_4 \mu_5})_{\alpha\beta}\nn\\
&&+Z^{[1] \mu}_{\mbox{\tiny NS}}(\tau_3)_{AB}(C\gamma_{\mu})_{\alpha\beta}
+\frac{1}{5!}Z^{[5(-)] \mu_1 \mu_2 \mu_3 \mu_4 \mu_5}_{\mbox{\tiny NS}}(\tau_3)_{AB}
(C\gamma_{\mu_1 \mu_2 \mu_3 \mu_4 \mu_5})_{\alpha\beta}\nn\\
&&+Z^{[0]}_{\mbox{\tiny RR}}(i\tau_2)_{AB}C_{\alpha\beta}
+\frac{1}{2!}Z^{[2] \mu \nu}_{\mbox{\tiny RR}}(\tau_1)_{AB}(C\gamma_{\mu \nu})_{\alpha\beta}\nn\\
&&+\frac{1}{4!}Z^{[4] \mu_1 \mu_2 \mu_3 \mu_4}_{\mbox{\tiny RR}}(i\tau_2)_{AB}
(C\gamma_{\mu_1 \mu_2 \mu_3 \mu_4})_{\alpha\beta}.\label{IIASUSY}
\eea
Here notation is followed in   
\cite{Hatsuda:1998by}
in which magnetic charges are not included yet. 
The number of  bosonic charges, $P_{\mu}$, $Z^{[5(+)]}_{\mbox{\tiny KK}}$,
$Z^{[1]}_{\mbox{\tiny NS}}$, 
$Z^{[5(-)]}_{\mbox{\tiny NS}}$,
$Z^{[0]}_{\mbox{\tiny RR}}$, 
$Z^{[2]}_{\mbox{\tiny RR}}$, 
$Z^{[4]}_{\mbox{\tiny RR}}$
is $(10+126)+(10+126)+(1+45+210)=528$.
Branes include 
a KK monopole with $Z^{[5(+)]}_{\mbox{\tiny KK}}$,
a fundamental string with $Z^{[1]}_{\mbox{\tiny NS}}$, 
a fundamental 5 brane with $Z^{[5(-)]}_{\mbox{\tiny NS}}$,
 D0 
 with $Z^{[0]}_{\mbox{\tiny RR}}$, 
 D2 and D8 with $Z^{[2]}_{\mbox{\tiny RR}}$ and 
 D4 and D6 with $Z^{[4]}_{\mbox{\tiny RR}}$.

\vskip 6mm

In $D=10$ the IIB theory has two 10-dimensional Majorana-Weyl supercharges 
which have the same chirality, 
$Q_{A\alpha}$ with $A=1,2$ and $\alpha=1,\cdots,16$.
The IIB superalgebra  is given by
\bea
\{Q_{A\alpha},Q_{B\beta}\}&=&
P_{\mu} \delta_{AB}(C\gamma^{\mu})_{\alpha\beta}
+\frac{1}{5!}Z^{[5+] \mu_1 \mu_2 \mu_3 \mu_4\mu_5}_{\mbox{\tiny KK}}\delta_{AB}
(C\gamma_{\mu_1 \mu_2 \mu_3 \mu_4 \mu_5})_{\alpha\beta}\nn\\
&&+Z^{[1] \mu}_{\mbox{\tiny NS}}(\tau_3)_{AB}(C\gamma_{\mu})_{\alpha\beta}
+\frac{1}{5!}Z^{[5+] \mu_1 \mu_2 \mu_3 \mu_4\mu_5}_{\mbox{\tiny NS}}(\tau_3)_{AB}
(C\gamma_{\mu_1 \mu_2 \mu_3 \mu_4 \mu_5})_{\alpha\beta}\nn\\
&&+Z^{[1] \mu}_{\mbox{\tiny RR}}(\tau_1)_{AB}(C\gamma_{\mu})_{\alpha\beta}
+\frac{1}{3!}Z^{[3] \mu \nu \rho}_{\mbox{\tiny RR}}(i\tau_2)_{AB}(C\gamma_{\mu \nu \rho})_{\alpha\beta}\nn\\&&
+\frac{1}{5!}Z^{[5+] \mu_1 \mu_2 \mu_3 \mu_4 \mu_5}_{\mbox{\tiny RR}}(\tau_1)_{AB}
(C\gamma_{\mu_1 \mu_2 \mu_3 \mu_4})_{\alpha\beta}.
\label{IIBSUSY}
\eea
The number of  bosonic charges, $P_m$, 
$Z^{[5+]}_{\mbox{\tiny KK}}$, $Z^{[1]}_{\mbox{\tiny NS}}$, 
$Z^{[5+]}_{\mbox{\tiny NS}}$,
$Z^{[1]}_{\mbox{\tiny RR}}$, 
$Z^{[3]}_{\mbox{\tiny RR}}$, 
$Z^{[5+]}_{\mbox{\tiny RR}}$
is $(10+126)+(10+126)+(10+120+126)=528$.
Branes include 
a KK monopole with $Z^{[5+]}_{\mbox{\tiny KK}}$,
a fundamental string with $Z^{[1]}_{\mbox{\tiny NS}}$, 
a fundamental 5 brane with $Z^{[5+]}_{\mbox{\tiny NS}}$,
 D1 and D9 with $Z^{[1]}_{\mbox{\tiny RR}}$, 
D3 and D7 with $Z^{[3]}_{\mbox{\tiny RR}}$ and 
 D5 with $Z^{[5+]}_{\mbox{\tiny RR}}$.

T-duality interchanges the IIA theory and the IIB theory.
Let us consider the T-duality transformation in the $x^9$ direction which is 
 interchanging the momentum $P_9$ and the winding mode $Z_{\mbox{\tiny NS}}^{[1]9}$.
In order to involve spinors the double space coordinates must be representations of the left and the right Lorentz groups, $P_{\rm L}{}^\mu=P^\mu+Z_{\mbox{\tiny NS}}^{[1]\mu}$,
$P_{\rm R}{}^\mu=P^\mu-Z_{\mbox{\tiny NS}}^{[1]\mu}$, $Q_{\rm L}=Q_1$ and $Q_{\rm R}=Q_2$.
Under the T-duality transformation double space coordinates are transformed as follows
\bea
\left(\begin{array}{c}P'_{\rm L}{}^{\mu'}\\P'_{\rm L}{}^9
	\\P'_{\rm R}{}^{\mu'}\\P'_{\rm R}{}^9\end{array}\right) 
=\left(\begin{array}{cc|cc}
	{\bf 1}&&&\\
	&1&&\\\hline
	&&{\bf 1}&\\&&&-1\end{array}\right)
\left(\begin{array}{c}P_{\rm L}{}^{\mu'}\\P_{\rm L}{}^9\\P_{\rm R}{}^{\mu'}\\P_{\rm R}{}^9
\end{array}\right) ~,~_{\mu'\neq 9}~.
\eea  
The $D=10, \mathcal{N}=2$ supercharges are transformed under the T-duality 
\bea
&(Q'_{\rm L},Q'_{\rm R})=(Q_{\rm L},Q_{\rm R})
\left(\begin{array}{cc}1&0\\0&\gamma^9
\end{array}\right)\Rightarrow
{\renewcommand{\arraystretch}{1.4}
\left\{\begin{array}{l}
\{Q_{\rm L},Q_{\rm L}\}=P_{\rm L}^\mu ~C\gamma_\mu +\cdots~\\
\{Q'{}_{\rm R},Q'{}_{\rm R} \}=P'{}_{\rm R}^\mu ~C\gamma_\mu +\cdots~\\
\{Q_{\rm L},Q'{}_{\rm R}\}=Z_{{\mbox{\tiny RR}};M} ~\Gamma^M\gamma^9
\end{array}\right.}
\eea
which changes the forms of the RR charges by $\gamma^9$.

The RR forms are transformed under the T-duality along $x^9$-direction summarized as below.
\begin{itemize}
{\item {\bf From IIA to IIB}
In addition to the interchanging the momentum and the winding mode 
the five forms are interchanged as
\bea
P_9\leftrightarrow Z_{\mbox{\tiny NS}}^{[1]9}~,~
Z^{[5(+)]9\mu_1'\mu_2'\mu_3'\mu_4'}_{\mbox{\tiny KK}}\leftrightarrow
Z_{\mbox{\tiny NS}}^{[5(-)]9\mu_1'\mu_2'\mu_3'\mu_4'}~,~_{\mu'\neq 9}~.
\eea
The IIB RR charges are read off from superalgebras, and they are written in terms of the IIA RR charges:
\bea
\begin{array}{ll}
	{\rm IIB}&	{\rm IIA}\\
	Z_{\mbox{\tiny RR}}^{[1]}& (Z_{\mbox{\tiny RR}}^{[2]},~Z_{\mbox{\tiny RR}}^{[0]})\\
	Z_{\mbox{\tiny RR}}^{[3]}&(Z_{\mbox{\tiny RR}}^{[4]},
	~Z_{\mbox{\tiny RR}}^{[2]})\\
	Z_{\mbox{\tiny RR}}^{[5+]}&
	Z_{\mbox{\tiny RR}}^{[4]}
		\end{array}
\eea 
}
{\item{\bf From IIB to IIA}

Analogously for the T-duality from the IIB to IIA the five forms are interchanged as
\bea
P_9\leftrightarrow Z_{\mbox{\tiny NS}}^{[1]9}~,~
Z^{[5+]9\mu_1'\mu_2'\mu_3'\mu_4'}_{\mbox{\tiny KK}}\leftrightarrow
Z_{\mbox{\tiny NS}}^{[5+]9\mu_1'\mu_2'\mu_3'\mu_4'}~,~_{\mu'\neq 9}~.
\eea
The IIA RR charges are written in terms of the IIB RR charges:
\bea
\begin{array}{ll}
	{\rm IIA}&	{\rm IIB}\\
	Z_{\mbox{\tiny RR}}^{[0]}& Z_{\mbox{\tiny RR}}^{[1]}\\
	Z_{\mbox{\tiny RR}}^{[2]}&(Z_{\mbox{\tiny RR}}^{[3]},
	~Z_{\mbox{\tiny RR}}^{[1]})\\
	Z_{\mbox{\tiny RR}}^{[4]}&
	(Z_{\mbox{\tiny RR}}^{[3]},~
	Z_{\mbox{\tiny RR}}^{[5+]})
\end{array}
\eea
}
\end{itemize}

\vskip 6mm

\subsection{Section conditions}

Now we go back to brane currents  $Z_M(\sigma)$ which are functions of 
 worldvolume coordinates.
They corresponds to 
the bosonic charges $Z_M$ of the IIA or IIB theories in \bref{IIASUSY} or \bref{IIBSUSY}.
Extended spaces are defined by  
current algebras generated by currents $Z_M(\sigma)$ as shown before.
In order to recover the physical 10-dimensional space 
we impose section conditions in the extended space.

\paragraph{$\mathcal{N} = (2,0)$ theory}

One finds that the worldvolume spatial diffeomorphism 
constraints are
translated into the vanishing condition of the following quantity
\cite{Hatsuda:2012uk, Hatsuda:2012vm, Hatsuda:2013dya}:
\begin{align}
Z_M \tilde{\rho}^{MN} Z_N = 0,~~~~_{M=(\mu,~ Y,~ \mu_1\mu_2,~ \mu_1Y,~ \mu_1\cdots \mu_5,~ \mu_1\cdots \mu_4 Y)} 
\label{eq:IIA_constraint}
\end{align}
where $\tilde{\rho}^{MN}$ is 
given by 
\begin{align}
  \left[
    \begin{array}{cccccc}
     0  &  0  &  a_{ [\nu_1 } \delta^{\mu}_{ \nu_2] }  &  0  &  b_{ [
      \nu_1 {\cdots} \nu_4 } \delta^{\mu}_{\nu_5]} & 0 \\
     0  &  0  &  0  &  a_{\nu_1}  &   0  &  b_{[ \nu_1 \cdots \nu_4
      ]}  \\
     a_{ [\mu_1 } \delta^{ \nu }_{ \mu_2] }  &  0  &  b_{ [\mu_1 \mu_2
      \nu_1 \nu_2] }  &  b_{ [\mu_1 \mu_2 \nu_1 Y] }  &  0  &  0  \\
     0  &  a_{\mu_1}  &  b_{ [\mu_1 \nu_1 \nu_2 Y] }  &  0  &  0  &
      0  \\
     b_{ [ \mu_1 \cdots \mu_4  } \delta^{\nu}_{\mu_5]} &  0  &   0  &
      0  &  0   &   0   \\
      0  &  b_{ [ \mu_1 \cdots \mu_4 ] } &  0  &  0  &  0  &  0
    \end{array}
  \right].
\end{align}
Here $a_{\mu}, b_{[\mu_1 \mu_2 \mu_3 \mu_4]}$ are arbitrary constants.

The constraint \eqref{eq:IIA_constraint} is expressed by the derivative
representations of the currents.
For the decomposition of the currents
\bea
Z_M=(Z_\mu,~Z^{[5(+)]\mu_1\mu_2\mu_3\mu_4\mu_5}_{\mbox{\tiny KK}};
~Z^\mu_{\mbox{\tiny NS}},~Z_{\mbox{\tiny NS}}^{[5(-)]\mu_1\mu_2\mu_3\mu_4\mu_5};
Z_{\mbox{\tiny RR}}^{[0]},~Z_{\mbox{\tiny RR}}^{[2]\mu_1\mu_2},
~Z_{\mbox{\tiny RR}}^{[4]\mu_1\mu_2\mu_3\mu_4}),
\eea
the coordinates in the extended space are denoted as
\bea
X^M=(x^\mu,~x_{\mu_1\mu_2\mu_3\mu_4\mu_5};
\tilde{x}_\mu,~\tilde{x}_{\mu_1\mu_2\mu_3\mu_4\mu_5};
~y,~y_{\mu_1\mu_2},~y_{\mu_1 \mu_2 \mu_3 \mu_4})~~~.
\eea
Then the Virasoro constraints 
\eqref{eq:IIA_constraint}
are explicitly written as 
\bea
{\renewcommand{\arraystretch}{1.6}\left\{\begin{array}{l}
		Z_\mu Z^{[1]\mu}_{\mbox{\tiny NS}}=0\\
		Z_\mu Z^{[2]\mu\nu}_{\mbox{\tiny RR}}+ Z_{\mbox{\tiny RR}}^{[0]} Z^{[1]\mu}_{\mbox{\tiny NS}} =0\\
		Z_\mu Z^{[5]\mu \nu_1\nu_2\nu_3\nu_4}
		+Z_{\mbox{\tiny RR}}^{[0]} Z^{[4]\nu_1\nu_2\nu_3\nu_4}_{\mbox{\tiny RR}}
		+\frac{1}{8}Z_{\mbox{\tiny RR}}^{[2][\nu_1\nu_2 }Z_{\mbox{\tiny RR}}^{[2] \nu_3\nu_4]}	=0\\
		Z_\mu Z_{\mbox{\tiny RR}}^{[4]\nu_1\nu_2\nu_3\nu_4}
		+\frac{1}{2}Z_{\mbox{\tiny RR}}^{[2][\nu_1\nu_2 }Z_{\mbox{\tiny NS}}^{[1] \nu_3]}=0
	\end{array}
	\right.}~~~,
\eea
The constraint \eqref{eq:IIA_constraint} is therefore given by the
following section conditions in the extended space:
\bea
{\renewcommand{\arraystretch}{1.6}\left\{\begin{array}{l}
		\partial_\mu\displaystyle\frac{\partial}{\partial \tilde{x}_{\mu}}=0\\	\partial_\mu\displaystyle\frac{\partial}{\partial y_{\mu\nu}}
		-\frac{\partial}{\partial y}
		\frac{\partial}{\partial \tilde{x}_{\nu}}  =0\\
		\partial_\mu\displaystyle\frac{\partial}{\partial x_{\mu\nu_1\nu_2\nu_3\nu_4}}+\frac{\partial}{\partial y}
		\frac{\partial}{\partial y_{\nu_1\nu_2\nu_3\nu_4}}
		+\frac{1}{8}\frac{\partial}{\partial y_{[\nu_1\nu_2}}
		\frac{\partial}{\partial y_{\nu_3\nu_4]}}
		=0\\
		\partial_\mu\displaystyle\frac{\partial}{\partial y_{\mu\nu_1\nu_2\nu_3\nu_4}}
		+\frac{1}{2}\frac{\partial}{\partial \tilde{x}_{[\nu_1}}
		\frac{\partial}{\partial y_{\nu_2\nu_3]}}
		=0
	\end{array}
	\right.}~~~.
\label{eq:section1}
\eea

We stress that the ``section condition'' \eqref{eq:section1} is defined
in the ten-dimensional (uncompactified) spacetime.
Therefore this provides a generalization of the section condition
\eqref{eq:section_condition_DFT} in DFT.
Indeed, when the target space is restricted to five-dimensions and the
index $\mu$ runs $1, \ldots, 5$, the matrix $\tilde{\rho}^{MN}$ 
in \eqref{eq:IIA_constraint} is expanded by 
$\tilde{\rho}^{MN} = \tilde{\rho}^{\underline{\mu} MN}
A_{\underline{\mu}}$ where $A_{\underline{\mu}} = (a_{\mu}, b_{[\mu_1
\cdots \mu_4]})$. 
In this case, one can show that $\tilde{\rho}^{\underline{\mu} MN}$ are
the $(5+5)$-dimensional gamma matrices and $\tilde{\rho}^{MN}$ becomes the
$SO(5,5)$ covariant tensor \cite{Hatsuda:2013dya}.
The condition \eqref{eq:section1} then reproduces the section condition
in the $SO(5,5)$ EFT in six dimensions \cite{Hohm:2013vpa}.

\paragraph{$\mathcal{N} = (1,1)$ theory}

The worldvolume diffeomorphism constraints ${\cal H}_i=0$
is written in a bilinear form by contracting with 
$E^i$,
$\epsilon^{i_1\cdots i_5} F_{i_1i_2}F_{i_3i_4} \partial_{i_5}x^\mu$
and \\
$\epsilon^{i_1\cdots i_5}F_{i_1i_2}\partial_{i_3}x^{\mu_1}\partial_{i_4}x^{\mu_4} \partial_{i_5}x^{\mu_3}$, 
$\epsilon^{i_1\cdots i_5}\partial_{i_1}x^{\mu_1}\cdots\partial_{i_5}x^{\mu_5}$,
\bea
&&~~
Z_M\tilde{\rho}^{MN}Z_N=0~,~~
~~{}_{M=(\mu, ~{\rm NS} \mu, ~{\rm RR} \mu, ~\mu_1\mu_2\mu_3,~ \mu_1\cdots \mu_5)} 
\nn\\~~&&
\tilde{\rho}^{MN}=
\left(
\begin{array}{cc|ccc}
	0&a\delta_{\nu}^{\mu}&b\delta^{\mu}_{\nu}&\cdots&
	c_{[5]}^{\mu}\\
	a\delta_{\mu}^{\nu}&0&\cdots&c_{[3]{\mu}}
	&0\\\hline
	b\delta^{\nu}_{\mu}&\vdots&&&\\
	\vdots&c_{[3]\nu}&&&\\
	c_{[5]}^{\nu}&0&&&
\end{array}
\right)
\nn\\
&&~~~~~c_{[5]}^{\mu}=\gamma_{[\nu_1\cdots \nu_{4}}\delta_{\nu_5]}^{\mu}~,~~~
~~~~~~c_{[3]\mu}=-_5C_2~\gamma_{[\nu_1\cdots \nu_{3} \mu]}
\eea
with arbitrary coefficients $a,~b,~\gamma$.
In concretely it is given as
\bea
\left\{{\renewcommand{\arraystretch}{1.6}
	\begin{array}{l}
p_{\mu} Z_{NS}^{[1]\mu}=0\\
p_{\mu} Z_{RR}^{[1]\mu}=0\\
p_{\mu} Z_{RR}^{[3]\mu \mu_1 \mu_2}+\frac{1}{2} Z_{NS}^{[1][\mu_1} Z_{RR}^{[1]\mu_2]}=0\\
p_{\mu} Z_{RR}^{[5]\mu \mu_1 \mu_2 \mu_3
 \mu_4} + 2 Z_{NS}^{[1] [\mu_1} Z_{RR}^{[3]\mu_2 \mu_3 \mu_4]}=0
\end{array}}
\right..
\eea
The section conditions in the extended space are given by
\bea
\left\{{\renewcommand{\arraystretch}{1.6}
	\begin{array}{l}
		\partial_{\mu} \displaystyle\frac{\del}{\del \tilde{x}_{\mu}} =0 \\
		\partial_{\mu} \displaystyle\frac{\del}{\del y_{\mu}} = 0 \\
		\partial_{\mu} \displaystyle\frac{\del}{\del y_{\mu \mu_1 \mu_2}}
		+\frac{1}{2} \displaystyle\frac{\del}{\del \tilde{x}_{[\mu_1}}
		\frac{\del}{\del \tilde{x}_{\mu_2]}} =0 \\
		\partial_{\mu} \displaystyle\frac{\del}{\del y_{\mu \mu_1 \mu_2 \mu_3
		 \mu_4}} + 2 \displaystyle\frac{\del}{\del \tilde{x}_{[\mu_1}}
		 \frac{\del}{\del y_{\mu_2 \mu_3 \mu_4]}} = 0
\end{array}}
\right..
\label{eq:section2}
\eea
Again, this is a generalization of the section condition
\eqref{eq:section_condition_DFT} in DFT.
As mentioned earlier, the condition \eqref{eq:section2}
is defined in the uncompactified, ten-dimensional target space.

\subsection{Gauge transformations}

Finally, we briefly comment on the gauge transformations of the type II
supergravity fields. 
As we have shown above, the current of the fundamental string generates
the gauge transformations of the spacetime vielbein and the $B$-field.
In the following, we derive the gauge transformations of type II 
supergravity fields by utilizing the current algebras of five-branes discussed so far.

\paragraph{$\mathcal{N} = (2,0)$ theory}
The gauge transformation of gauge fields $E_A{}^M$ with $E_A{}^ME_B{}^N\hat{\eta}^{AB}={\cal M}^{MN}$ and
$E_A{}^ME_B{}^N{\eta}^{AB}=\eta^{MN}$ 
is given as follows.
The type IIA gauge fields are represented as a vector in the extended
space 
where the index of a vector $V^M$ runs 
$M=({}^\mu,
~{}_\mu,~{}_{\cdot},~{}_{\mu_1\mu_2},~{}_{\mu_1,\cdots,\mu_4})$
\bea
E_a{}^M=(e_a{}^\mu,~e_a{}^\nu B_{\nu\mu},~e_a{}^\nu {C}^{[1]}_{\nu},~e_a{}^{\nu}{C}^{[3]}_{\nu\mu_1\mu_2},~e_a{}^\nu{C}^{[5]}_{\nu\mu_1\mu_2\mu_3\mu_4})~~.
\eea
The $D$-dimensional vielbein $e_a{}^\mu$ is invertible.
The gauge parameter vector is given as 
\bea
\lambda^M=(\lambda^\mu,~\lambda_{\mu},~\lambda^{[0]},~\lambda^{[2]}_{\mu_1\mu_2},~\lambda^{[4]}_{\mu_1\mu_2\mu_3\mu_4})~~~.
\eea
The gauge transformation rules are given by the commutators of the gauge field vector $E_a{}^MZ_M$ and the parameter vector $\lambda^MZ_M$ in the extended space as
\bea
\delta_\lambda E_a{}^M Z_M(\sigma)&=&
\im \int d\sigma' \{ \lambda^NZ_N(\sigma'), E_a{}^MZ_M(\sigma)\}~~~.
\eea
For this computation the section conditions are used, then the worldvolume coordinate derivatives are calculated by the usual chain rule, $\partial_i\Phi(x^\mu)=\partial_i x^\mu \partial_\mu \Phi$ resulting the closure of the gauge transformation. 

The transformation rules are calculated as
\bea
\left\{
{\renewcommand{\arraystretch}{1.6}
	\begin{array}{l}
		\delta_\lambda e_a{}^\mu={\cal L}_\lambda e_a{}^\mu=\lambda^\nu\partial_\nu e_a{}^\mu-
		e_a{}^\nu\partial_\nu\lambda^\mu\\
		\delta_\lambda B_{\nu\mu}={\cal L}_\lambda B_{\nu\mu}+\partial_{[\mu}\lambda_{\nu]}\\
		\delta_\lambda {C}^{[1]}_{\nu\mu}={\cal L}_\lambda {C}^{[1]}_{\nu\mu}+\partial_{\mu}\lambda^{[0]}\\
		\delta_\lambda {C}^{[3]}_{\mu_1\mu_2\mu_3}={\cal L}_\lambda {C}^{[3]}_{\mu_1\mu_2\mu_3\mu_4}+  \frac{1}{3!}\partial_{[\mu_1}\lambda^{[2]}_{\mu_2\mu_3]}
			+\frac{1}{3!}B_{[\mu_1\mu_2}\partial_{\mu_3]}\lambda^{[0]}\\
		\delta_\lambda {C}^{[5]}_{\mu_1\mu_2\mu_3\mu_4\mu_5}
		={\cal L}_\lambda{C}^{[5]}_{\mu_1\mu_2\mu_3\mu_4\mu_5}
		+\frac{1}{5!}\partial_{[\mu_1}\lambda^{[4]}_{\mu_2\mu_3\mu_4\mu_5]}\\
		\quad\quad\quad\quad\quad\quad\quad\quad\quad+\frac{2}{5!}B_{[\mu_1\mu_2}\partial_{\mu_3}\lambda^{[2]}_{\mu_4\mu_5]}
		+\frac{2}{5!}{C}^{[3]}{}_{[\mu_1\mu_2\mu_3}\partial_{\mu_4}\lambda_{\mu_5]}
\end{array}}
\right..
\eea

\paragraph{$\mathcal{N} = (1,1)$ theory}

The type IIB gauge fields are a represented as a vector in the extended space 
where the index of a vector $V^M$ runs 
	$M=({}^\mu,
	~_{\rm NS}{}_\mu,~_{\rm RR}{}_{\mu},~{}_{\mu_1\mu_2\mu_3},~{}_{\mu_1,\cdots,\mu_5})$
\bea
E_a{}^M=(e_a{}^\mu,~e_a{}^\nu B_{\nu\mu},~e_a{}^\nu{C}^{[2]}_{\nu\mu},~e_a{}^\nu{C}^{[4]}_{\nu\mu_1\mu_2\mu_3},~e_a{}^\nu{C}^{[6]}_{\nu\mu_1\mu_2\mu_3\mu_4\mu_5}) ~~~.
\eea
The gauge parameter vector is given as 
\bea
\lambda^M=(\lambda^\mu,~\lambda_{\mu},~\lambda^{[1]}_{\mu},~\lambda^{[3]}_{\mu_1\mu_2\mu_3},~\lambda^{[5]}_{\mu_1\mu_2\mu_3\mu_4\mu_5})~~~.
\eea
The transformation rules are calculated as
\bea
\left\{
{\renewcommand{\arraystretch}{1.6}
	\begin{array}{l}
		\delta_\lambda e_a{}^\mu={\cal L}_\lambda e_a{}^\mu\\
		\delta_\lambda B_{\nu\mu}={\cal L}_\lambda B_{\nu\mu}+\partial_{[\mu}\lambda_{\nu]}\\
		\delta_\lambda {C}^{[2]}_{\nu\mu}={\cal L}_\lambda {C}^{[2]}_{\nu\mu}+\partial_{[\mu}\lambda^{[1]}_{\nu]}\\
		\delta_\lambda {C}^{[4]}_{\mu_1\mu_2\mu_3\mu_4}={\cal L}_\lambda {C}^{[4]}_{\mu_1\mu_2\mu_3\mu_4}+\frac{1}{3!}\partial_{[\mu_1}\lambda^{[3]}_{\mu_2\mu_3\mu_4]}
		+\frac{1}{3!}B_{[\mu_1\mu_2}\partial_{\mu_3}\lambda_{\mu_4]}
		+\frac{1}{3!}B_{[\mu_1\mu_2}\partial_{\mu_3}\lambda^{[1]}_{\mu_4]}\\
		\delta_\lambda {C}^{[6]}_{\mu_1\mu_2\mu_3\mu_4\mu_5\mu_6}
		={\cal L}_\lambda{C}^{[6]}_{\mu_1\mu_2\mu_3\mu_4\mu_5\mu_6}
		+\frac{1}{5!}\partial_{[\mu_1}\lambda^{[5]}_{\mu_2\mu_3\mu_4\mu_5\mu_6]}\\
		\quad\quad\quad\quad\quad\quad\quad\quad\quad+\frac{2}{6!}B_{[\mu_1\mu_2}\partial_{\mu_3}\lambda^{[3]}_{\mu_4\mu_5\mu_6]}
		+\frac{2}{6!}{C}^{[4]}{}_{[\mu_1\mu_2\mu_3\mu_4}\partial_{\mu_5}\lambda_{\mu_6]}
\end{array}}
\right..
\eea

\section{Conclusion and discussions} \label{sect:conclusion}

\begin{figure}[t]
\begin{center}
\includegraphics[scale=.8]{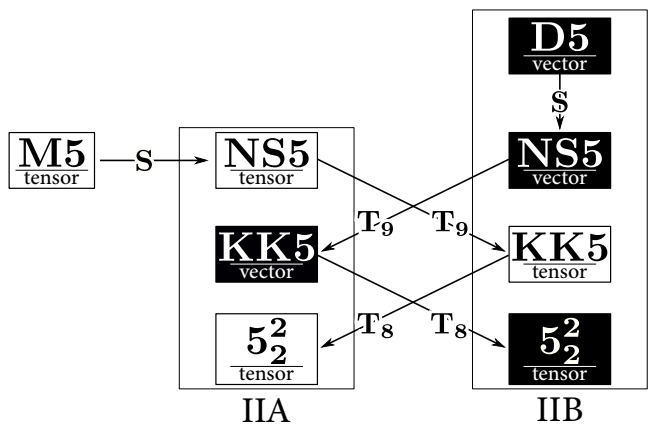}
\end{center}
\caption{Type II five-branes related by S- and T-dualities.}
\label{fig:duality}
\end{figure}

In this paper, we studied the current algebras of five-branes in type II
string theories.
These include the NS5-branes, the KK5-branes and the exotic
$5^2_2$-branes in type IIA/IIB string theories.
Together with the M2-, M5-branes, fundamental strings and the
D$p$-branes, they share important roles in string theories. 
The effective theories of the five-branes are characterized by the
six-dimensional $\mathcal{N} = (2,0)$ tensor and the $\mathcal{N} = (1,1)$
vector multiplets.
The worldvolume effective actions of the $\mathcal{N} = (2,0)$ tensor
multiplet have been derived from the PST action of the M5-brane
by the double dimensional reduction and the T-duality transformations.
The effective actions of the $\mathcal{N} = (1,1)$ vector multiplet have
been derived from the DBI action of the D5-brane by the S- and T-duality
transformations.
See Figure \ref{fig:duality}.

The actions of the KK5-branes and the $5^2_2$-branes contain geometric
fluctuation modes $x^{\mu}$ together with the dual scalar modes
$\tilde{x}^a$ and the gauge fields in the tensor and the vector
multiplets.
Indeed, the algebras of the KK5- and the $5^2_2$-branes have specific
fluctuation modes $\tilde{x}^a \, (a=8,9)$ corresponding to the isometry
directions in the transverse space.
We explicitly wrote down the current algebras of the five-branes based
on these actions.
The existence of these dual scalar modes in the KK5- and the
$5^2_2$-branes distinguish their algebras from those of the D$p$-branes.
Due to the self-duality constraint of the 2-form gauge field, 
we showed that algebras of the $\mathcal{N} = (2,0)$ theories are given
by the Dirac bracket. 
On the other hand, the algebras of the $\mathcal{N} = (1,1)$ theories
are given by the Poisson bracket.

The currents $Z_M$ in the algebras are naturally expressed
by the differential operators in the extended spaces.
The coordinates $X^M$ in the extended spaces are the fundamental basis
of the doubled field theory (DFT) and the exceptional field theory
(EFT).
They include the winding coordinates of various branes.
We show that the worldvolume spatial diffeomorphism constraints result
in the physical conditions in the extended spaces.
The diffeomorphism constraints of five-branes and the corresponding
section conditions \eqref{eq:section1}, \eqref{eq:section2} are 
defined in the uncompactified ten-dimensional spacetime.
Therefore, in a sense, they can be recognized as a general expressions of the section
conditions for EFT in lower-dimensions.
We note that the section conditions found in this paper are represented in the 528-dimensional extended space.
Whether this extended space fits into the $E_{n(n)}$ languages or not deserves further studies.

Based on the current algebras, we also wrote down the Courant and Dorfman brackets for the gauge
parameters in ten-dimensional type II supergravities.
We then derive the gauge transformations of the supergravity fields in type
IIA and IIB theories.
We also discuss the relation between the currents and the central
charges in the ten-dimensional supersymmetry algebras. 

We have presented the current algebras of the five-branes including the
exotic $5^2_2$-branes.
However, we note that they are just the entrance to the web of exotic branes.
The U-duality multiplets of the 1/2 BPS objects in the $T^d$ compactification have
been classified by the duality group $E_{d(d)}$.
There are a large number of exotic branes whose tensions are
proportional to $g_s^{-\alpha}$ with $\alpha \ge 2$
\cite{Fernandez-Melgarejo:2018yxq, Berman:2018okd}.
Branes whose tensions are proportional to $g_s^{-3}, g_s^{-4}, \ldots$ are still mysterious objects.
However, this is not the end of the story.
There are yet unfamiliar exotic objects in string theory.
One can perform further T-duality transformations along the transverse
directions to the $5^2_2$-brane worldvolumes.
The resulting objects are denote as $5^3_2$- and $5^4_2$-branes and known
as the R-branes and space-filling branes \cite{Hassler:2013wsa}.
Since their geometries are characterized, not only by the geometrical
coordinate $x^{\mu}$ but also the T-dualized one $\tilde{x}_{\mu}$, they
are not solutions to conventional supergravities.
This kind of branes are called locally non-geometric objects.
They are shown to be solutions to DFT, EFT and deformed
supergravities \cite{Berman:2014jsa, Bakhmatov:2016kfn, Kimura:2018hph, Berman:2018okd,
Fernandez-Melgarejo:2018yxq} 
and their worldvolume theories are discussed
\cite{Shiozawa:2020inc}.
It is known that the locally non-geometric nature appear in various
context of string theory \cite{Gregory:1997te, Harvey:2005ab, Kimura:2013fda,
Kimura:2013zva, Lust:2017jox, Kimura:2018hph, Kimura:2018ain}.
It would be interesting to study the current algebras of these branes. 
We will come back to these issues in future studies.

\subsection*{Acknowledgments}
The work of S.S. is supported by Grant-in-Aid for Scientific Research (C), JSPS KAKENHI Grant Number JP20K03952.



\end{document}